\documentclass[a4paper,11pt]{article}
\pdfoutput=1 

\usepackage{jcappub} 

\usepackage[T1]{fontenc} 
\usepackage{mathtools}

\usepackage{enumitem}

\usepackage{aas_macros}
\usepackage{amsmath,amssymb}
\usepackage{multirow}
\usepackage{cleveref}
\usepackage{siunitx}
\usepackage{enumerate}
\usepackage[dvipsnames]{xcolor}
\usepackage{float}

\usepackage[compat=1.1.0]{tikz-feynman}
\usepackage{afterpage}

\usepackage{amsthm}

\usepackage{psfrag}
\usepackage{color}
\usepackage{graphicx}
\usepackage{upgreek}
\usepackage{feynmp}
\DeclareMathAlphabet{\mathpzc}{OT1}{pzc}{m}{it}

\usepackage[mathscr]{euscript}

\usepackage{jcappub} 
\usepackage[T1]{fontenc} 
\usepackage[toc,page]{appendix}
\usepackage{circledsteps}
\usepackage{natbib}
\setcitestyle{square,comma,numbers,sort&compress}
\usepackage{enumerate}
\usepackage{tabularx,booktabs}
\usepackage[left = 1.0in, top=1.0in,right=1.0in,bottom=1.0in,a4paper]{geometry}
\usepackage[dvipsnames]{xcolor}
\usepackage{comment}
\usepackage[caption=false]{subfig}
\usepackage[compat=1.1.0]{tikz-feynman} 
\def\mc#1{\mathcal#1}
\usepackage[section]{placeins}
\usepackage{cleveref}

\allowdisplaybreaks[4]

\definecolor{darkgreen}{rgb}{0,0.5,0}

\newcommand{\beq}{\begin{eqnarray}}
\newcommand{\eeq}{\end{eqnarray}}

\newcommand{\bseq}{\begin{subequations}}
\newcommand{\eseq}{\end{subequations}}
\newcommand{\be}{\begin{equation}}
\newcommand{\ee}{\end{equation}}

\renewcommand{\Im}{\mathop{\rm Im}\nolimits}

\newcommand{\beqa}{\begin{eqnarray}}
\newcommand{\eeqa}{\end{eqnarray}}

\usepackage{simpler-wick}
\newcolumntype{Y}{>{\centering\arraybackslash}X}

\title{Parity-Odd and Even Trispectrum from Axion Inflation}

\author[a]{Xuce Niu,}
\author[a,b]{Moinul Hossain Rahat,}
\author[a]{Karthik Srinivasan,}
\author[a]{and 
\quad \quad \quad 
Wei Xue}

\affiliation[a]{
Department of Physics,
University of Florida, Gainesville, FL 32611, USA }
\affiliation[b]{School of Physics \& Astronomy, University of Southampton, Southampton SO17 1BJ, UK }
\emailAdd{xuce.niu@ufl.edu}
\emailAdd{M.H.Rahat@soton.ac.uk}
\emailAdd{karthik.srinivas@ufl.edu}
\emailAdd{weixue@ufl.edu}

\abstract{
The four-point correlation function of primordial scalar perturbations has parity-even and parity-odd 
contributions and the parity-odd signal in cosmological observations is opening a novel window to look for
new physics in the inflationary epoch. We study the distinct parity-odd and even prediction from the axion inflation 
model, in which the inflaton couples to a vector field via a Chern-Simons interaction, and 
the vector field is considered to be either approximately massless ($m_A \ll $ Hubble scale $H$) or very massive ($m_A \sim H $).
The parity-odd signal arises due to one transverse mode of the vector field being predominantly produced during inflation. 
We adopt the in-in formalism to evaluate the correlation functions. 
Considering the vector field mode function to be dominated by its real part up to a constant phase,
we simplify the formulas for numerical computations. The numerical studies show that the massive and massless vector fields
give significant parity-even signals, while the parity-odd contribution is about one to two orders of magnitude smaller.
}
\begin{document}
\maketitle
\flushbottom

\section{Introduction}
\label{sec:intro}

We are in the new age of precision cosmology with a large influx of cosmological data. The current and future experiments 
give us an excellent opportunity to understand the fundamental physics during the period of inflation \cite{Brout:1977ix, Starobinsky:1980te, 
Kazanas:1980tx, Sato:1980yn, Guth:1980zm, Linde:1981mu, Albrecht:1982wi, Linde:1983gd}. 
The information of the early universe is encoded in primordial perturbations, including scalar and tensor fluctuations, and can be
observed in the cosmic microwave background (CMB),
the large-scale structure (LSS)
and gravitational waves.

The primordial perturbations generated during cosmic inflation lead to observational consequences, which have been widely investigated 
through the fluctuations' two- and three-point correlation functions to constrain inflationary scenarios and 
probe signatures of new physics. Inflation predicts Gaussian perturbations with minor deviations. The magnitude and shape of the
non-Gaussianities will help us pin down the models of inflation and understand the inflaton kinetic terms \cite{Chen:2006nt, Cheung:2007st}, potential \cite{Adshead:2011jq}, and 
vacuum \cite{Chen:2006nt, Holman:2007na,Meerburg:2009ys,Ashoorioon:2010xg}. Furthermore, the feature of non-Gaussianities may probe new particles during inflation.
Observing oscillatory shape in high-point correlation function could be an unambiguous way to detect the presence of new particles during 
inflation -- a program that has been named ``cosmological collider physics'' 
\cite{Chen:2009zp, Chen:2009we, Baumann:2011nk, Arkani-Hamed:2015bza, Chen:2016nrs, 
Lee:2016vti, Meerburg:2016zdz, Chen:2016uwp, Chen:2016hrz,An:2017hlx,Kumar:2017ecc, Chen:2018xck,Wu:2018lmx,Li:2019ves, 
Lu:2019tjj,Hook:2019zxa, Hook:2019vcn, Kumar:2019ebj, Wang:2019gbi,Wang:2020uic, Wang:2020ioa, 
Maru:2021ezc, Lu:2021gso, Wang:2021qez, Tong:2021wai, Cui:2021iie, Pinol:2021aun, Tong:2022cdz, Reece:2022soh, Jazayeri:2022kjy, 
Pimentel:2022fsc, Chen:2022vzh, Qin:2022lva, Maru:2022bhr}. 

The four-point correlation function, known as trispectrum, is more difficult to detect due to the noise backgrounds but provides
further information about inflaton and its interactions. The trispectrum prediction from single field inflation is studied in \cite{Seery:2006vu,Arroja:2008ga,Chen:2006dfn,Seery:2008ax}.
For the cosmological collider signatures, the oscillatory features from the four-point functions have been investigated in \cite{Arkani-Hamed:2015bza}. 
In addition to the prediction for the magnitude and shape of the trispectrum, the four-point correlation function 
conveys the smallest statistics for probing parity violation. 
Parity invariance implies that the physics of a system remains unaltered with the reversal of coordinate axes. 
For scalar perturbations, parity-odd signatures can only be observed in four- or higher-point correlation functions, 
since, for a coplanar configuration of objects in three dimensions, a parity transformation is equivalent to a rotation.

In this paper we 
investigate a scenario where a pseudoscalar 
inflaton $\phi$ couples to a $U(1)$ gauge boson through a Chern-Simons term $\phi F \tilde{F}$, 
where $F$ is the field-strength of the gauge field and $\tilde{F}$ is its dual. 
This class of inflation is referred to as axion inflation, and the axion inflation model and its 
phenomenology has been extensively studied in \cite{Domcke:2019mnd, Adshead:2012kp, Adshead:2015pva, Adshead:2016iae, Adshead:2018doq, Adshead:2019lbr, Adshead:2019igv, Freese:1990rb, Silverstein:2008sg, McAllister:2008hb, Kim:2004rp, Berg:2009tg, Dimopoulos:2005ac, Pajer:2013fsa}. 
Its coupling to the $U(1)$ gauge boson leads to copious production of a transverse gauge mode, whose inverse decay into the inflaton sources additional primordial fluctuations, 
leading to a parity-violating tensor power spectrum and a large bispectrum at CMB scales \cite{ Anber:2006xt,Anber:2009ua,Cook:2011hg, Barnaby:2010vf, Barnaby:2011qe, Barnaby:2011vw, Meerburg:2012id, Anber:2012du, Linde:2012bt, Cheng:2015oqa, Garcia-Bellido:2016dkw, Domcke:2016bkh, Peloso:2016gqs, Domcke:2018eki, Cuissa:2018oiw,Sorbo:2011rz,Odintsov:2022hxu,Nojiri:2020pqr,Nojiri:2019nar,Odintsov:2019mlf,Cho:2019plw}.
The chirality of the resultant gravitational wave spectrum can be detected by combining observations from multiple future interferometers 
\cite{Orlando:2020oko}. Further, we focus on the four-point correlation function of the curvature perturbations from axion inflation,
motivated by the recent observation of parity-odd galaxy trispectrum from 
the BOSS galaxy data \cite{Cahn:2021ltp, Hou:2022wfj,Philcox:2022hkh} and the cosmological collider physics.      
The four-point correlation function appears only at the loop level in the axion inflation model. The parity odd signals from different models and 
tree-level contribution are studied in \cite{Shiraishi:2016mok,Liu:2019fag,Cabass:2022rhr,Cabass:2022oap}. 
We explore either the massless or massive gauge boson. The massless case applies to $m_A = 0$ and $m_A \ll H$, where $H$ is the Hubble scale of inflation. 
It covers a wide range of parameter space in this model. The massive gauge boson $m_A \sim H$ study is particularly interesting and 
has unique imprints from various observational consequences at the cosmological colliders \cite{Wang:2020ioa,Niu:2022quw}.

The paper is organized as follows. In \cref{sec:axioninflation}, we review the dynamics of gauge mode production in axion inflation. \Cref{sec:correlationfunc} contains our analysis of the $1$-loop four-point correlation function and discusses the real mode function approximation to simplify the calculations. We present our numerical results in \cref{sec:results} and conclude in \cref{sec:conclusion}.

\section{Axion inflation dynamics}\label{sec:axioninflation}
We consider a single field slow-roll inflation, where the inflaton $\phi$ is an axion-like scalar with an approximate shift symmetry 
$\phi \to \phi +  {\rm const}$. 
Due to the shift symmetry, the inflaton $\phi$ is naturally coupled to a $U(1)$ 
gauge field $A_\mu$ via a Chern-Simons operator, $- \phi F \tilde{F} / (4 \Lambda)$. So the action takes the form, 
\begin{align}
    S = \int d^4x \sqrt{-g} \left[   
        - \frac{1}{4}F^{\mu\nu}F_{\mu\nu}+\frac{1}{2}m_A^2 A^\mu A_\mu - 
         \frac{1}{4\Lambda}\phi \Tilde{F}^{\mu\nu}F_{\mu\nu} \right], \label{action}
\end{align}
where $F_{\mu\nu} \equiv \partial_\mu A_\nu -\partial_\nu A_\mu$ is the field strength of the gauge field, 
and $\Tilde{F}^{\mu\nu} \equiv \frac{1}{2} \frac{\epsilon^{\mu\nu\alpha\beta}}{\sqrt{-g}}F_{\alpha\beta}$ is its dual.
The spacetime background is quasi-de Sitter space with the metric given as    
\begin{align}
    ds^2 \equiv g_{\mu\nu}dx^\mu dx^\nu 
      = dt^2 - a^2(t) dx_i dx^i 
   = a^2(\tau) (d\tau^2 - \delta_{ij} dx^i dx^j),
\end{align}
where $t$ is the physical time and $\tau$ is the conformal time.
The scale factor $a (\tau ) \simeq - 1 / ( H \tau )$ and $H$ as the Hubble scale slowly varies in time. 

We consider massive and massless gauge boson production due to the rolling inflaton. 
For the massive gauge boson, we take the mass about the Hubble scale $m_A \sim H$; for the massless gauge boson, we take 
$m_A= 0$ or its mass is much lighter than the Hubble scale $m_A \ll H$ and is approximately to zero in the inflationary background.  
The massive vector field with the constraints $\partial_\mu (\sqrt{-g}A^\mu) = 0$ 
is decomposed using the helicity basis $\lambda = \pm , 0$, 
\begin{align}
    \mathbf{A}(\tau, \mathbf{x}) = \sum_{\lambda = \pm,0} \int \frac{d^3 k}{(2\pi)^3} 
      \left[ \boldsymbol{\epsilon}_\lambda (\mathbf{k}) a_\lambda (\mathbf{k}) A_\lambda (\tau, k) 
         e^{i \mathbf{k}\cdot \mathbf{x}} + \text{h.c.} \right] \label{gaugedecomp}
\end{align}
where $ \boldsymbol{\epsilon}_\lambda (\mathbf{k})$ is the polarization vector\footnote{For a $3$-vector momentum $\mathbf{k} \equiv k(\sin{\theta}
\cos{\phi}, \sin{\theta}\sin{\phi}, \cos{\theta})$, the transverse polarization vector is defined as
 $   \boldsymbol{\epsilon_\pm}(\mathbf{k}) \equiv \frac{1}{\sqrt{2}} \left(\mp \cos{\theta} \cos{\phi} + i\sin{\phi}, \mp \cos{\theta} \sin{\phi} - i\cos{\phi}, \pm \sin{\theta} \right)
$.} $a_\lambda (\mathbf{k})$ is the annihilation 
operator, and $A_\lambda (\tau, k)$ is the wavefunction solution.   
The dominant vector field production is governed by the field equations of the transverse modes,
\begin{align}
   \partial_\tau^2{{A}_\pm} ( \tau, k) + \left(k^2 + {a(\tau)^2} m_A^2 \pm \frac{2k\xi}{\tau}\right) {{A}_\pm} ( \tau, k)  &= 0,
   \label{eq:Aeq}
\end{align}
and $\xi$ is defined as 
\begin{equation}
   \xi \equiv \frac{\dot{\phi}_0}{2\Lambda H} \, .
\end{equation}
The rolling of the inflaton background field $\dot \phi_0 = \frac{d \phi_0(t) } {d t} $ 
is taken to be constant as a good approximation during inflation,
and $\dot \phi_0 > 0$ is assumed here so that $A_+$ mode has instability and is dominantly produced. Because we concentrate on
a transverse mode, the massless vector field results are the same as the massive one by setting $m_A =0$ in solutions.
Satisfying the initial condition of the Bunch-Davies vacuum, we find the solution of mode function as
\begin{equation}
    {A}_\pm (\tau, k ) =  
     \frac{1} {\sqrt{2k}} e^{\pm\pi \xi/2} W_{\mp i\xi, i\mu}(2ik\tau), \label{Whittaker}
\end{equation}
where $W$ stands for the Whittaker function, 
and the parameter $\mu \equiv \sqrt{(m_A/H)^2 - 1/4}$. Note that adding an overall phase $e^{i \theta_0}$ does not change any physical results. 
For the massless vector field, we can use the mode function solution in \cref{Whittaker},
or express the solutions by the Coulomb wave functions \cite{Anber:2009ua}, 
\begin{equation}
    {A}_\pm (\tau, k )  = \frac{1} {\sqrt{2k}} \left[ G_0 ( \xi, - k \tau) + i F_0 (\xi , - k\tau) \right] \, .
    \label{eq:Am0}
\end{equation}
The two solutions in \cref{Whittaker,eq:Am0} are identical for $m_A= 0 $ up to a constant phase term.

For the scalar field, we decompose it into an unperturbed term $\phi_0(t)$ and its perturbation $\delta \phi(t, {\bf x} )$,
\begin{equation}
   \phi (t, {\bf x}) = \phi_0(t) + \delta \phi(t, {\bf x} ) \, .
\end{equation}
We assume that the backreaction of gauge field into the inflaton background evolution is negligible, though this effect is essential towards
the end of inflation \cite{Barnaby:2011qe,Linde:2012bt}. 
However, the gauge field can significantly change the small inflaton perturbations via the Chern-Simons coupling, so that it gives a sizable loop
corrections to the correlation functions of curvature perturbations.

\section{Parity-odd and parity-even correlation functions}
\label{sec:correlationfunc}

This section will discuss the four-point correlation functions of curvature perturbations, 
which are closely related
to the observables in the CMB and the LSS. The four-point correlation functions are composed of parity 
even and parity odd parts. The in-in formalism \cite{Weinberg:2005vy} is introduced to evaluate the correlation function. 
For the inflaton coupling to vector fields via the Chern-Simons terms, a significant simplification is achieved when 
considering the predominant real wavefunction solutions of the vector fields at the cost of losing some accuracy.

\subsection{Correlation functions of curvature perturbation}

The curvature perturbation $\zeta( t, {\bf x} )$
is originated in the quantum fluctuations during the period of inflation, 
is a conserved quantity after being stretched outside the Hubble horizon for single-field inflation, 
and predicts the late universe's scalar modes of density and temperature perturbations. 
We adopt the convention in Ref.~\cite{Maldacena:2002vr}. The curvature perturbation $\zeta( t, {\bf x} )$ is given in the spatial part of metric 
when choosing the gauge that the inflation perturbation is zero. The metric takes the form  $g_{ij} = a^2 e^{2 \zeta } [ \exp \gamma ]_{ij} $, 
where $\gamma_{ij}$ is the tensor fluctuations. The curvature perturbation is related to the inflaton fluctuations in the 
spatial flat gauge, 
\begin{equation}
   \zeta( t, {\bf x} ) =  - \frac {H} { \dot \phi_0} \delta \phi (t, {\bf x} )  \, .
\label{eq:phi2zeta}
\end{equation}
To consider the Chern-Simons interaction in the axion inflation, the spatial flat gauge is more convenient and thus is used here.

We expand the curvature perturbation in the Fourier space, 
\begin{equation}
   \zeta( t, {\bf x} )  = \int  \frac{d^3 k}{(2\pi)^3} \zeta ( t, {\bf k } ) \, e^{i {\bf k} \cdot {\bf x} } \, .  
\end{equation}
The reality of $ \zeta( t, {\bf x} )$ leads to the condition of $  \zeta^* ( t, {\bf k } )   =  \zeta ( t,  - {\bf k } ) $.
Under the Parity transformation that ${\bf x} \to - {\bf x}$ and ${\bf k } \to - {\bf k}$, the n-point correlation function at a fixed
time is transformed as
\begin{equation}
   {\rm Parity:} \quad \left\langle \textstyle \prod_i^n   \zeta ( t, {\bf k }_i ) \right\rangle  \rightarrow   
         \left\langle {\textstyle \prod}_i^n   \zeta ( t, - {\bf k }_i ) \right\rangle
         = \left\langle {\textstyle \prod}_i^n   \zeta ( t,  {\bf k }_i ) \right\rangle^*
\end{equation}
Therefore, the real part of the correlation function is parity even, while the imaginary part will change the signatures under the 
transformation so that it is parity odd. The parity odd or the imaginary part arises for four- and higher-point correlation
functions, which may be understood by the rotation symmetry. For the three-point correlation function, the three momenta 
${\bf k}_i$ are in the same plane
due to the momentum conservation, and after parity transformation, the 2d configuration can be converted into itself with a 3d rotation.
Alternatively, we express the three-point function as a function of three independent parameters 
\begin{equation}
    \langle \zeta ( t, {\bf k }_1 )   \zeta ( t, {\bf k }_2 )  \zeta ( t, {\bf k }_3 )
      \rangle^\prime  = G^{(3)} \left( k_1, k_2 ,  {\bf k }_1  \cdot {\bf k}_2 \right)  \, ,
\end{equation}
where prime denotes the correlation function stripped off the $\delta$-function, $(2\pi)^3 \delta^3( \textstyle \sum_i {\bf k}_i)$.
The three parameters are invariant under parity transformation, so that the three-point function should be real and parity even.
But for the four-point correlation function, some extra parameters are introduced, such as, 
${\bf k}_1 \cdot ({\bf k}_2 \times {\bf k}_3 ) $, which is parity odd. Hence four-point correlation function can have parity odd signatures.

The four-point function, also called trispectrum, takes the form 
\begin{equation}
    \langle \zeta ( t, {\bf k }_1 )   \zeta ( t, {\bf k }_2 )  \zeta ( t, {\bf k }_3 )
     \zeta ( t, {\bf k }_4 )    \rangle =   
          (2\pi )^{9} \delta^3 ( {\bf k}_1 + {\bf k}_2 + {\bf k}_3 + {\bf k}_4)  \,  
     P_\zeta^3  \, \frac{\textstyle \sum_i  k_i^3 } { \textstyle \prod_i  k_i^3 }  
    {\cal T} ({\bf k}_1, {\bf k}_2,  {\bf k}_3, {\bf k}_4) 
    \label{eq:4pointTff}
\end{equation}
where ${\cal T} ({\bf k}_1, {\bf k}_2,  {\bf k}_3, {\bf k}_4)$ is the dimensionless form factor, is also denoted as ${\cal T}$ for simplicity.    
The power spectrum of the scalar fluctuations is given by the two-point correlation function at the tree level, and 
is evaluated by the inflaton fluctuations $\delta \phi $ and the relation of 
\cref{eq:phi2zeta},
\begin{equation}
     \langle \zeta ( t, {\bf k }_1 )   \zeta ( t, {\bf k }_2 )  \rangle_{(0)}  =  
     (2\pi )^{5} \delta^3 ( {\bf k}_1 + {\bf k}_2 ) \frac{1}{ 2k^3}
     P_\zeta ( k ) \, , 
     \quad
    P_\zeta ( k )  = \frac{ H^2} { \dot \phi_0^2 }  \frac{ H^2} { ( 2\pi)^2 } \, .
    \label{eq:Pzeta}
\end{equation}
where we assign $k_1 = k_2 =k$.
At the leading order, the power spectrum is $k$ independent.

\subsection{In-in formalism and the correlation function from gauge field production}

We employ the in-in formalism \cite{Weinberg:2005vy} as a proper tool to evaluate the correlation functions. For a general operator 
at a fixed time ${\cal O} (\tau) $, the conditions of the fields are imposed at the early time in the inflation epoch and the in-in formalism 
takes the form
\begin{equation}
   \langle {\cal O} ( \tau ) \rangle
   = \left\langle
      \left[ {\bar T} \exp ( i \int_{-\infty}^\tau d\tau H_I (\tau ) )  \right]
    {\cal O}_I ( \tau )
      \left[ { T} \exp ( - i \int_{-\infty}^\tau d\tau H_I (\tau ) )  \right]
   \right\rangle  \ ,
   \label{eq:inin1}
\end{equation}
where
$T$ is a time-ordered product and $\bar{T}$ is an anti-time-ordered product. $H_I (\tau )$ is the interaction part of the Hamiltonian in
the interaction picture and ${\cal O}_I$ denotes the operator in the interaction picture as well.
We adopt an equivalent and more convenient formula than \cref{eq:inin1} presented in Ref.~\cite{Weinberg:2005vy}, 
\begin{equation}
   \begin{split}
   \langle {\cal O} ( \tau ) \rangle
   &=
   \sum_{N= 0}^{\infty} i^N \int_{-\infty}^0 d\tau_N \int_{-\infty}^{\tau_N} d\tau_{N-1} \cdots  \int_{-\infty}^{\tau_2}  d\tau_1
      \, 
      \langle 
      [ H_I ( \tau_1),  \cdots [H_I (\tau_N), {\cal O}_I ( \tau ) ] \cdots ] 
      \rangle 
   \, .
   \end{split}
   \label{eq:inin2}
\end{equation}
Further, we consider the current interaction, $H_I (\tau ) = - \int d^3x  \delta \phi J $,
take the assumption that the current $J$ does not contain $\delta \phi$ field and $\delta \phi$ does not show in the internal line.
This interaction form and assumption are valid for the axion inflation. In the axion inflation,
the gauge field production may significantly change the curvature perturbations and impact the n-point correlation 
functions with $n > 2$, through the inflaton coupling with a current,   
\begin{equation}
   S_J = \int d\tau d^3{ x} \, \delta\phi ( \tau, {\bf x} ) \, J ( \tau,  {\bf x} )
       = \int d\tau \frac{d^3{k}}{(2\pi)^3} \, \delta\phi ( \tau, -{\bf k} ) \, J_{\bf k} ( \tau  )
         \ .
\end{equation}
The source takes the form
\begin{equation}
  J ( \tau,  {\bf x} ) = - \frac{1}{ 8 \Lambda}  \epsilon^{\mu \nu \alpha \beta } F_{\mu \nu } F_{\alpha \beta}
\end{equation}
and its Fourier transform is given as
\begin{equation}
   J_{\mathbf{k}}(\tau)  =  \frac{a^4(\tau)}{\Lambda} \int d^3x e^{-i \mathbf{k}\cdot \mathbf{x}} \mathbf{E} \cdot \mathbf{B} \label{Jk} \, .
\end{equation}
where $\mathbf{E}$ and $\mathbf{B}$ are the electric and magnetic fields associated with the vector field,
\begin{align}
    \mathbf{E} = -\frac{1}{a^2} \mathbf{A'}, \qquad \mathbf{B} = \frac{1}{a^2} \boldsymbol{\nabla} \times \mathbf{A}. \label{electromagnetic}
\end{align}
and $\mathbf{A'} = \partial_\tau \mathbf{A}$.
The measurement is at a fixed time $\tau_0 = 0$, after the end of inflation, and the classical value of the inflaton perturbation,
\begin{equation}
   \delta\phi_k ( \tau )   =  \frac{ H  }  { \sqrt{ 2 k^3} } ( 1 + i k \tau ) e^{ - i k \tau } \, ,
\end{equation}
becomes real, $\delta\phi_k ( 0  )  =  { H  }  / { \sqrt{ 2 k^3} } $.

\subsubsection*{2-point function}
The vector field production gives the two-point correlation function of curvature fluctuations a one-loop radiative correction,
\begin{equation}
   \begin{split}
    \langle  \zeta ( \tau_0 , {\bf k}_1 )   \zeta ( \tau_0 , {\bf k}_2 ) \rangle_{(1)}
       &=  i^2 \left( - \frac{H}{\dot \phi_0 } \right)^2
       \delta\phi_{k}^2(0 )
      \times 2 \, 
        \int_{-\infty}^0 d\tau_2  \int_{-\infty}^{\tau_2} d\tau_1 
       \left(  \delta\phi_k (\tau _2)
        -   \delta\phi_k^* (\tau _2)  \right)
      \\
      & 
    \left[
      \delta\phi_k( \tau_1 )
       \langle
       J_{{\bf k}_1} ( \tau_1) 
       J_{{\bf k}_2} ( \tau_2 ) 
       \rangle
      -
      \delta\phi_k^*( \tau_1 )
       \langle
       J_{{\bf k}_2} ( \tau_2) 
       J_{{\bf k}_1} ( \tau_1 ) 
       \rangle
       \right]
   \end{split}
   \label{eq:zetazeta}
\end{equation}
Considering the momentum conservation from the current correlation function 
$ \langle J_{{\bf k}_1} ( \tau_1) J_{{\bf k}_2} ( \tau_2 ) \rangle$, we assign $k_1= k_2 = k$.
The current correlation function should sum over all the individual modes of the vector field, but the dominant one is
the $A_+$, and we thus consider the $A_+$ mode only, 
\begin{equation} 
   \begin{split}
    \langle
       J_{{\bf k}_1} ( \tau_1) 
       J_{{\bf k}_2} ( \tau_2 ) 
    \rangle
   =& \frac{ ( -1)^2 }{ 2   \Lambda^2} ( 2\pi )^3 \delta^3 ( {\bf k}_1+ {\bf k}_2 )   
      \int\frac{ d^3 q_1}{ ( 2\pi )^3} 
         | \boldsymbol{\epsilon}_+ ( {\bf q}_1 )  \cdot 
          \boldsymbol{\epsilon}_+ ( -  {\bf q}_2 )  |^2 
   \\
   & \times 
      {\cal B}_2 ( q_1, q_2 , \tau_1 ) 
      {\cal B}_2 ( \underline{q_2 } , \underline{ q_1 }, \tau_2 ) 
   \end{split}
   \label{eq:JJ}
\end{equation} 
where 
\begin{equation} 
    {\bf q}_2   = {\bf q}_1   - {\bf k}_1 \, , 
\end{equation} 
$1/2$ is a symmetry factor for counting the equivalent contractions of the vector fields, and 
the condition of the polarization $ \boldsymbol{\epsilon}_+^* ( {\bf p}) = \boldsymbol{\epsilon}_+ ( -{\bf p} )$ is used.
The contractions lead to the function ${\cal B}_2 $,  
\begin{subequations}
   \begin{align}
      {\cal B}_2 ( q_1 , q_2, \tau ) &\equiv q_1  A_+(\tau,q_1) A_+'(\tau,q_2) + q_2 A_+(\tau, q_2 )
         A_+'(\tau,q_1)  \, , 
      \\
      {\cal B}_2 ( \underline{q_1} , q_2, \tau ) &\equiv q_1  A_+^*(\tau,q_1) A_+'(\tau,q_2) + q_2 A_+(\tau, q_2 )
         A_+'^*(\tau,q_1)  \, . 
   \end{align}
\end{subequations}
An underline on the loop momentum on the left-hand side corresponds to a complex conjugation of the associated mode functions. 
(and their derivatives). For $ {\cal B}_2 ( \underline{q_2 }, \underline{ q_1 }, \tau_2 ) $, we take the complex conjugate of the two mode functions.  
Therefore, putting together \cref{eq:zetazeta,eq:JJ}, we have the two-point correction function from the one-loop radiative correction,
\begin{equation}
   \begin{split}
    \langle  \zeta ( \tau_0 , {\bf k}_1 )   \zeta ( \tau_0 , {\bf k}_2 ) \rangle_{(1)}^\prime
       &=  -  \left( - \frac{H}{\dot \phi_0 } \right)^2 \frac{H^2} { 2 k^3} \frac{1}{\Lambda^2}
        \int_{-\infty}^0 d\tau_2  \int_{\infty}^{\tau_2} d\tau_1 
      \int\frac{ d^3 q_1}{ ( 2\pi )^3} 
         | \boldsymbol{\epsilon}_+ ( {\bf q}_1 )  \cdot 
          \boldsymbol{\epsilon}_+ ( - {\bf q}_2  )  |^2 
      \\
      & 
    \left[
      \delta\phi_k( \tau_1 )
       (  \delta\phi_k (\tau _2)
        -   \delta\phi_k^* (\tau _2)  )
      {\cal B}_2 ( q_1 , q_2, \tau_1 ) 
      {\cal B}_2 ( \underline{q_2 } , \underline{ q_1 }, \tau_2 ) 
       + \rm{c.c.}
       \right],
   \end{split}
\end{equation}
where c.c. stands for the complex conjugate.
Since the expression inside the square bracket is real, it is evident that the two-point function is parity even.

\subsubsection*{3-point function}
For completeness, we also present the three-point correlation function result,
\begin{eqnarray}
    && \langle  \zeta ( \tau_0 , {\bf k}_1 )   \zeta ( \tau_0 , {\bf k}_2 ) 
         \zeta ( \tau_0 , {\bf k}_3 )   \rangle^\prime_{(1)}
      \nonumber
      \\
       &=&     
      i^3 \left( - \frac{H}{\dot \phi_0 } \right)^3   \frac{H^3} { ( 2  k_1 k_2 k_3   )^{3/2}} \frac{1}{\Lambda^3}
        \int_{-\infty}^0 d\tau_3 
        \int_{-\infty}^{\tau_3} d\tau_2  \int_{-\infty}^{\tau_2} d\tau_1 
      \nonumber
      \\
      && 
       \int\frac{ d^3 q_1}{ ( 2\pi )^3} 
          \boldsymbol{\epsilon}_+ ( {\bf q}_1 )  \cdot 
          \boldsymbol{\epsilon}_+ ( - {\bf q}_2  )   \, 
          \boldsymbol{\epsilon}_+ ( {\bf q}_2 )  \cdot 
          \boldsymbol{\epsilon}_+ ( - {\bf q}_3  )  \, 
          \boldsymbol{\epsilon}_+ ( {\bf q}_3 )  \cdot 
          \boldsymbol{\epsilon}_+ ( - {\bf q}_1  )  \,  
      \nonumber
      \\
      &&
    \bigg[
      \delta\phi_{k_1}( \tau_1 )
      \delta\phi_{k_2}( \tau_2 )
     (  \delta\phi_{k_3}( \tau_3 )
      -   \delta\phi_{k_3}^*( \tau_3 ) )
      {\cal B}_2 ( q_1 , q_2, \tau_1 ) 
      {\cal B}_2 ( \underline{q_2 } , q_3 ,  \tau_2 ) 
      {\cal B}_2 (  \underline{q_3} , \underline{q_1 } ,  \tau_3 ) 
      \nonumber
      \\
      &&
      - \delta\phi_{k_1}( \tau_1 )
      \delta\phi_{k_2}^*( \tau_2 )
     (  \delta\phi_{k_3}( \tau_3 )
      -   \delta\phi_{k_3}^*( \tau_3 ) )
      {\cal B}_2 ( q_1 , q_2, \tau_1 ) 
      {\cal B}_2 ( {q_3 } ,\underline{q_1 },  \tau_3 ) 
      {\cal B}_2 (  \underline{q_2} , \underline{q_3 } ,  \tau_2 ) 
      \nonumber
      \\
      &&
      +  {\rm 5 \, perms }
      - \rm{c.c.}
       \bigg],
\end{eqnarray}
where ${\bf q}_1$, ${\bf q}_2$ and ${\bf q}_3$ relation is given by the momentum conservation of the three vertices, 
\begin{equation}
   {\bf q}_2  = {\bf q}_1  - {\bf k}_1 \, , 
      \quad
   {\bf q}_3  = {\bf q}_1  + {\bf k}_1 \, .
   \label{eq:q23}
\end{equation}
The rule to determine the momentum flow direction is that the momentum of ${\bf q}_i$ flows into the vertex having the external field 
of momentum ${\bf k}_i$. The $+ {\rm 5 \, perms}$ take account of the exchange of $( {\bf k}_1 , {\bf k}_2, {\bf k}_3 )$.   
The permutations would change the definition of ${\bf q}_i$ in \cref{eq:q23}.\footnote{Alternatively, since ${q}_i$ is a dummy variable, 
if we fix the definition of ${\bf q}_i$ in \cref{eq:q23}, then the permutation should also apply to the function ${\cal B}_2$.
For the product of three ${\cal B}_2$, we perform the permutation of the three pair $( q_1, q_2) $ , $( q_2, q_3) $ and $( q_3, q_1) $ 
to account for the three vertex exchanges. The complex conjugate of the mode function in ${\cal B}_2$ needs to be modified accordingly.}
The product of $i^3$ and the terms in the square bracket is real due to the $- \rm{c.c.}$ part, so that the imaginary part of the three-point function may
be given by the polarization vector. However, the imaginary part of the polarization terms is proportional to 
${\bf k}_1 \cdot ( {\bf k}_2 \times {\bf k}_3 ) = 0 $. The three-point function only provides parity-even signals.

\subsubsection*{4-point function}

\begin{figure}[!ht]
    \centering
    \subfloat[$F1$\label{F1diagram}]{
        \includegraphics[width=0.25\textwidth]{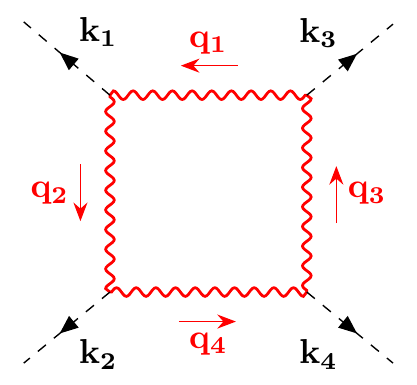}
    } \hspace*{0.5cm}
    \subfloat[$F2$\label{F2diagram}]{
        \includegraphics[width=0.25\textwidth]{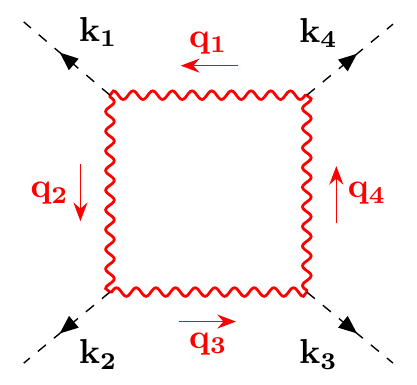}
    }
    \hspace*{0.5cm}
    \subfloat[$F3$\label{F3diagram}]{
        \includegraphics[width=0.25\textwidth]{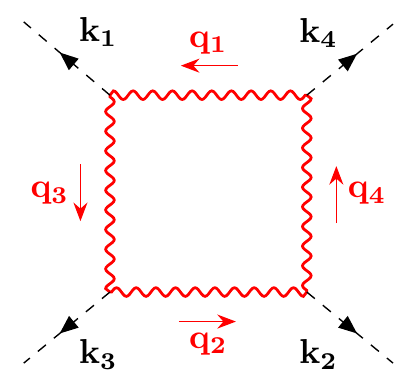}
    }
    \caption{Three independent diagrams for four-point function.}
    \label{fig:Feynman}
\end{figure}

The calculation of the four-point function is similar to the three-point function, while the results contain three independent diagrams shown in 
\cref{fig:Feynman}. The correlation function of the first diagram takes the form 
\begin{eqnarray}
    && \langle  \zeta ( \tau_0 , {\bf k}_1 )   \zeta ( \tau_0 , {\bf k}_2 ) 
         \zeta ( \tau_0 , {\bf k}_3 )  \zeta ( \tau_0 , {\bf k}_4 )  \rangle_{(F1)}^\prime
      \nonumber
      \\
       &=&    i^4  \left( - \frac{H}{\dot \phi_0 } \right)^4   \frac{H^4} { 4 (k_1 k_2 k_3 k_4)^{3/2}}  \frac{1}{\Lambda^4}
        \int_{-\infty}^0 d\tau_4  \int_{\infty}^{\tau_4} d\tau_3
        \int_{-\infty}^{\tau_3} d\tau_2  \int_{-\infty}^{\tau_2} d\tau_1 
       \int\frac{ d^3 q_1}{ ( 2\pi )^3} 
      \nonumber
      \\
      && 
          \boldsymbol{\epsilon}_+ ( {\bf q}_1 )  \cdot 
          \boldsymbol{\epsilon}_+ ( - {\bf q}_2  )   \, 
          \boldsymbol{\epsilon}_+ ( {\bf q}_2 )  \cdot 
          \boldsymbol{\epsilon}_+ ( - {\bf q}_4  )  \, 
          \boldsymbol{\epsilon}_+ ( {\bf q}_4 )  \cdot 
          \boldsymbol{\epsilon}_+ ( - {\bf q}_3  )  \,  
          \boldsymbol{\epsilon}_+ ( {\bf q}_3 )  \cdot 
          \boldsymbol{\epsilon}_+ ( - {\bf q}_1  )  \,  
      \nonumber
      \\
      &&
    \bigg[
      \delta\phi_{k_1}( \tau_1 )
      \delta\phi_{k_2}( \tau_2 )
      \delta\phi_{k_3}( \tau_3 )
     (  \delta\phi_{k_4}( \tau_4 )
      -   \delta\phi_{k_4}^*( \tau_4 ) )
      {\cal B}_2 ( q_1 , q_2, \tau_1 ) 
      {\cal B}_2 ( \underline{q_2 } , q_4 ,  \tau_2 ) 
      {\cal B}_2 (  q_3 , \underline{q_1 } ,  \tau_3 ) 
      {\cal B}_2 (  \underline{q_4} , \underline{q_3 } ,  \tau_4 ) 
      \nonumber
      \\
      &&
      - \delta\phi_{k_1}( \tau_1 )
      \delta\phi_{k_2}( \tau_2 )
      \delta\phi_{k_3}^*( \tau_3 )
     (  \delta\phi_{k_4}( \tau_4 )
      -   \delta\phi_{k_4}^*( \tau_4 ) )
      {\cal B}_2 ( q_1 , q_2, \tau_1 ) 
      {\cal B}_2 ( \underline{q_2 } , q_4 ,  \tau_2 ) 
      {\cal B}_2 (  \underline{q_4} , q_3  ,  \tau_4 ) 
      {\cal B}_2 (  \underline{q_3} , \underline{q_1 } ,  \tau_3 ) 
      \nonumber
      \\
      &&
      - \delta\phi_{k_1}( \tau_1 )
      \delta\phi_{k_2}^*( \tau_2 )
      \delta\phi_{k_3}( \tau_3 )
     (  \delta\phi_{k_4}( \tau_4 )
      -   \delta\phi_{k_4}^*( \tau_4 ) )
      {\cal B}_2 ( q_1 , q_2, \tau_1 ) 
      {\cal B}_2 (  {q_3} , \underline{q_1 } ,  \tau_3 ) 
      {\cal B}_2 (  {q_4} , \underline{q_3}  ,  \tau_4 ) 
      {\cal B}_2 ( \underline{q_2 } , \underline{q_4} ,  \tau_2 ) 
      \nonumber
      \\
      &&
      + \delta\phi_{k_1}( \tau_1 )
      \delta\phi_{k_2}^*( \tau_2 )
      \delta\phi_{k_3}^*( \tau_3 )
     (  \delta\phi_{k_4}( \tau_4 )
      -   \delta\phi_{k_4}^*( \tau_4 ) )
      {\cal B}_2 ( q_1 , q_2, \tau_1 ) 
      {\cal B}_2 (  {q_4} , {q_3}  ,  \tau_4 ) 
      {\cal B}_2 (  \underline {q_3} , \underline{q_1 } ,  \tau_3 ) 
      {\cal B}_2 ( \underline{q_2 } , \underline{q_4} ,  \tau_2 ) 
      \nonumber
      \\
      &&
      +  {\rm 23 \, perms }
      + \rm{c.c.}
       \bigg] \, .
   \label{eq:4pointF1}
\end{eqnarray}
The correlation functions from the other two diagrams are presented in \cref{app:4point}.
As shown in the first diagram of \cref{fig:Feynman}, the relation of the momentum ${\bf q}_i$ is obtained by the momentum conservation at each 
vertex,
\begin{equation}
   {\bf q}_2  = {\bf q}_1  - {\bf k}_1 \, , 
      \quad
   {\bf q}_3  = {\bf q}_1  + {\bf k}_1 \, ,
      \quad
   {\bf q}_4  = {\bf q}_1  - {\bf k}_1 - {\bf k}_2 \, .
   \label{eq:q234}
\end{equation}
There are $24$ permutations of the four momentum vectors $({\bf k}_1, {\bf k}_2, {\bf k}_3, {\bf k}_4 )$. The $c.c.$ takes the real value of 
the terms in the square bracket. Hence, the parity-odd part of the four-point function is given by the polarization vectors. Compared with the three-point function,  
the imaginary part is also proportional to ${\bf k}_1 \cdot ( {\bf k}_2 \times {\bf k}_3 ) $,
but in this case, the four-momentum vectors are not necessary for the same plane. The imaginary part could be nonzero, which is confirmed in numerical studies.

\subsection{Real mode function approximation} 
\label{subsec:dominantreal}

The general formulas of the four-point function in eqs.~\eqref{eq:4pointF1}, \eqref{eq:4pointF2} and \eqref{eq:4pointF3}
containing ${\cal O} (100)$ terms plus 7-dimensional integration are complicated for numerical calculations. A great simplification is achieved
when the real part of the gauge boson's mode function $A_+$ up to a constant phase is dominant in the time domain when
the gauge boson production effects are essential. Here we will derive the simplified correlation function formulas, 
which have a similar form in Ref.~\cite{Barnaby:2011vw}. The dominance of the real part of the mode functions for the massless and massive vector bosons
will be justified numerically. 

We consider a current interaction $H_I = - \int d^3 x  \phi J $ and assume that the mode function of $A_+$ in \cref{Whittaker} is 
real up to an unphysical overall 
constant phase $e^{i \theta_0}$. This assumption implies that the mode function is classical and 
is valid when $- k \tau \lesssim {\cal O}(1)$, one gauge boson mode is predominantly produced. 
Under this assumption, ${\cal B}_2 (q_i,q_j, \tau ) = {\cal B}_2 ( \underline{q_i},q_j, \tau ) = {\cal B}_2 ( {q_i},\underline{q_j}, \tau ) =
{\cal B}_2 ( \underline{q_i},\underline{q_j}, \tau )  $, such that the four-point function in eq.~(\ref{eq:4pointF1}) simplifies to 
\begin{eqnarray}
    && \langle  \zeta ( \tau_0 , {\bf k}_1 )   \zeta ( \tau_0 , {\bf k}_2 ) 
         \zeta ( \tau_0 , {\bf k}_3 )  \zeta ( \tau_0 , {\bf k}_4 )  \rangle_{(F1)}^\prime
       =     \left(  \frac{H}{\dot \phi_0 } \right)^4   \frac{2^4 H^4} { 4 (k_1 k_2 k_3 k_4)^{3/2}}  \frac{1}{\Lambda^4}
      \nonumber
      \\
      && \times 
       \int\frac{ d^3 q_1}{ ( 2\pi )^3} 
          \boldsymbol{\epsilon}_+ ( {\bf q}_1 )  \cdot 
          \boldsymbol{\epsilon}_+ ( - {\bf q}_2  )   \, 
          \boldsymbol{\epsilon}_+ ( {\bf q}_2 )  \cdot 
          \boldsymbol{\epsilon}_+ ( - {\bf q}_4  )  \, 
          \boldsymbol{\epsilon}_+ ( {\bf q}_4 )  \cdot 
          \boldsymbol{\epsilon}_+ ( - {\bf q}_3  )  \,  
          \boldsymbol{\epsilon}_+ ( {\bf q}_3 )  \cdot 
          \boldsymbol{\epsilon}_+ ( -{\bf q}_1 )  
      \nonumber
      \\
      &&
      \times 
       \int_{-\infty}^{0} d\tau_1 \Im \delta\phi_{k_1}( \tau_1 ) {\cal B}_2 ( q_1 , q_2, \tau_1 )  \, 
       \int_{-\infty}^{0} d\tau_2 \Im \delta\phi_{k_2}( \tau_2 ) {\cal B}_2 ( \underline{q_2} , q_4, \tau_2 ) 
      \nonumber
      \\
      &&
      \times 
       \int_{-\infty}^{0} d\tau_3 \Im \delta\phi_{k_3}( \tau_3 ) {\cal B}_2 ( q_3 , \underline{q_1}, \tau_3 ) 
       \int_{-\infty}^{0} d\tau_4 \Im \delta\phi_{k_4}( \tau_4 ) {\cal B}_2 ( \underline{q_4} , \underline{q_3}, \tau_4 ) 
      \, .
   \label{eq:sourceF1}
\end{eqnarray}
We keep the underline of momentum in ${\cal B}_2$ to note that the overall constant phase is irrelevant. 
A simpler approach to derive this formula may start from the general in-in formalism in \cref{eq:inin1} and use the fact 
that the time-ordering and anti-time-ordering do not influence the current with only the real mode function. 
The time integration in eq.~(\ref{eq:4pointF1}) reduces
to four independent time integration in eq.~\eqref{eq:sourceF1}. 
Similarly, the simplified formulas for the second and third diagrams are presented in \cref{app:4point}.

\begin{figure}[t]
\begin{center}
 \includegraphics[width=0.45\textwidth]{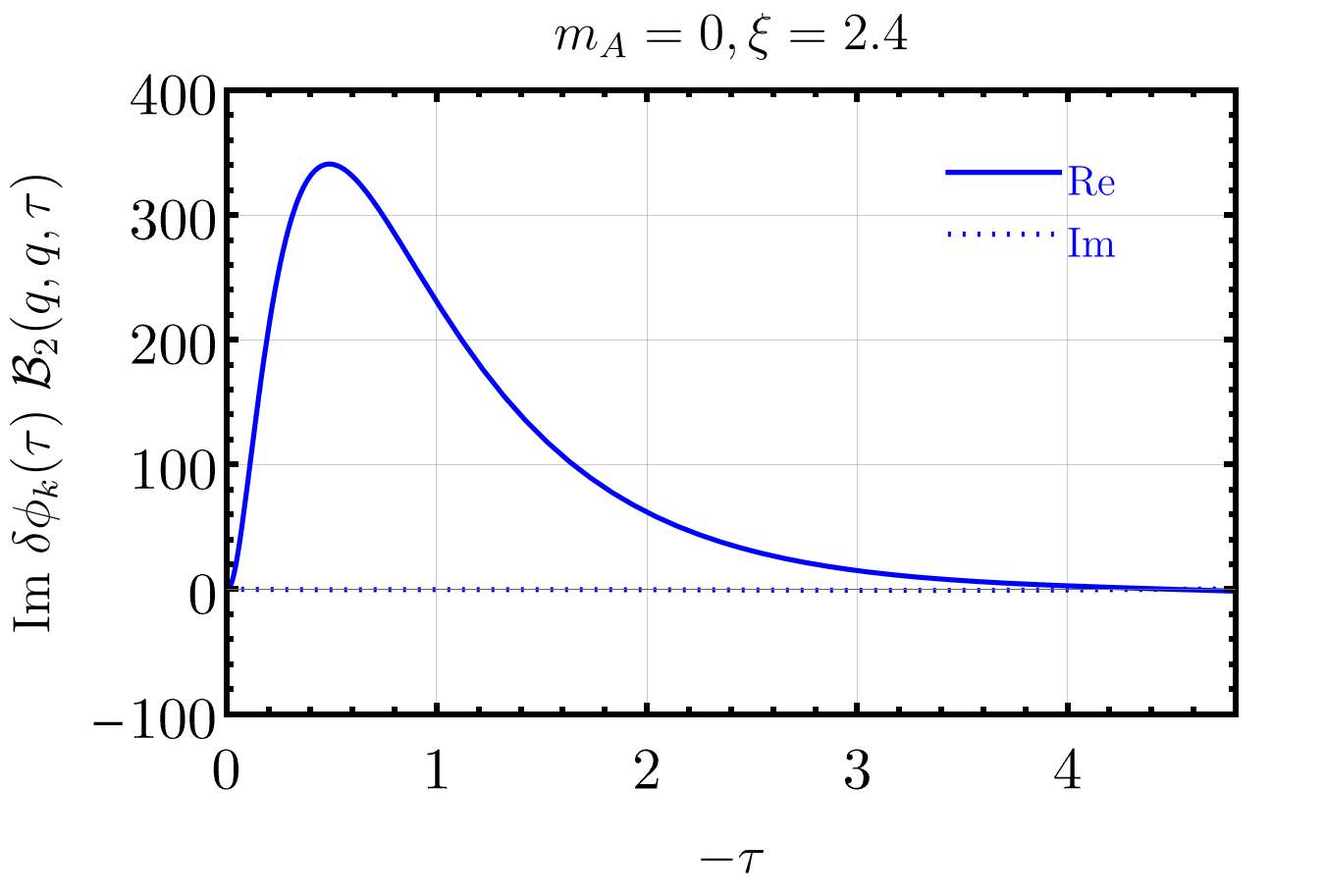}
\qquad
\quad
 \includegraphics[width=0.45\textwidth]{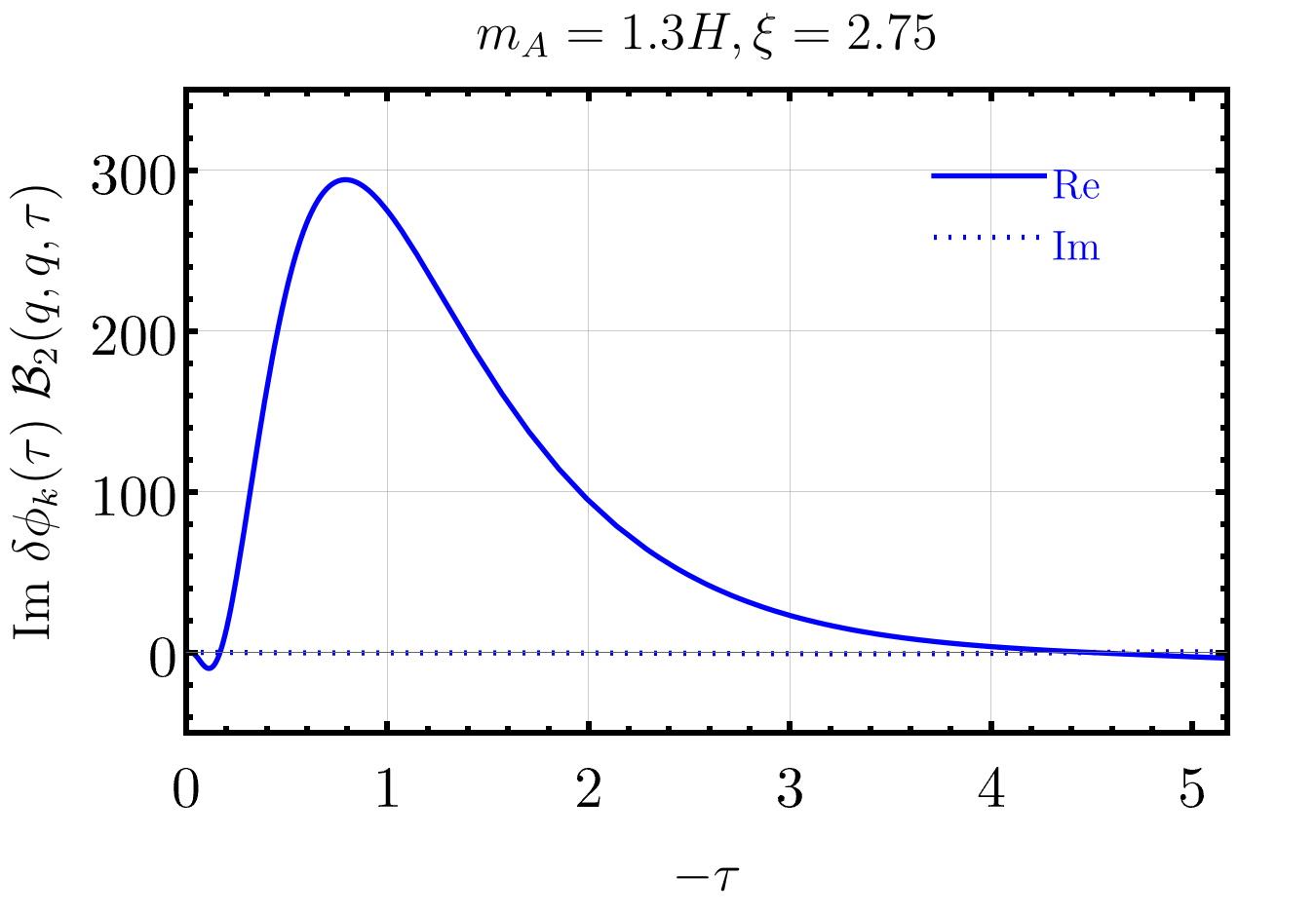}
\caption{The integrand $\Im \delta \phi_k ( \tau ) {\cal B}_2 ( q, q , \tau)$ for massless and massive vector boson.
The real part of the function shows in the solid curves, and the imaginary one is given in the dotted curves.
In the massless case, we choose the parameter $\xi = 2.4$; in the massive case, we choose $m = 1.3$ and $\xi = 2.75$.
In both plots, we set $H=1$, $k=1$ and $q=1$ and vary the time $\tau$. 
}
\label{fig:integrand}
\end{center}
\end{figure}

\begin{figure}[t]
\begin{center}
 \includegraphics[width=0.44\textwidth]{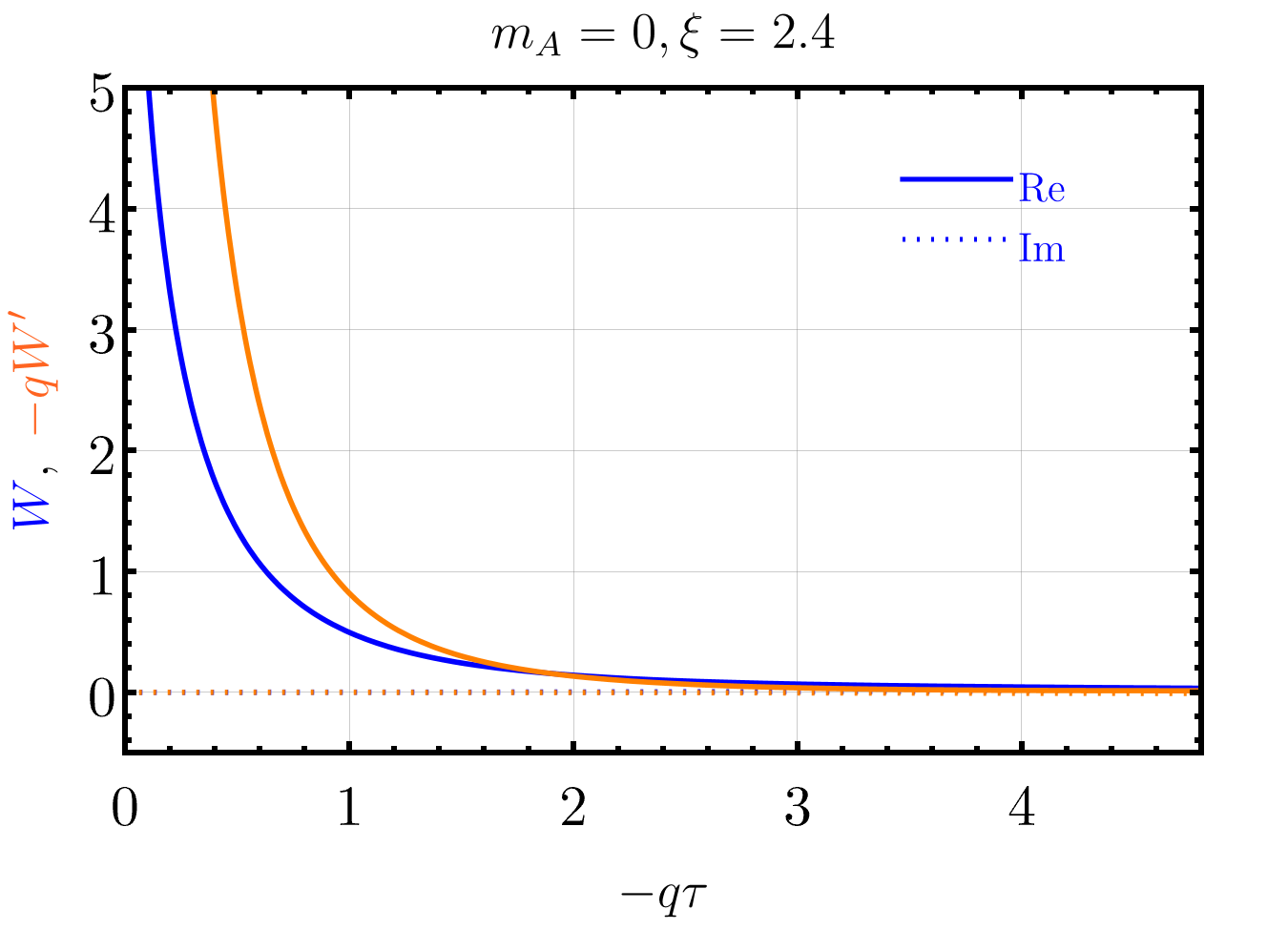}
\qquad
\quad
 \includegraphics[width=0.47\textwidth]{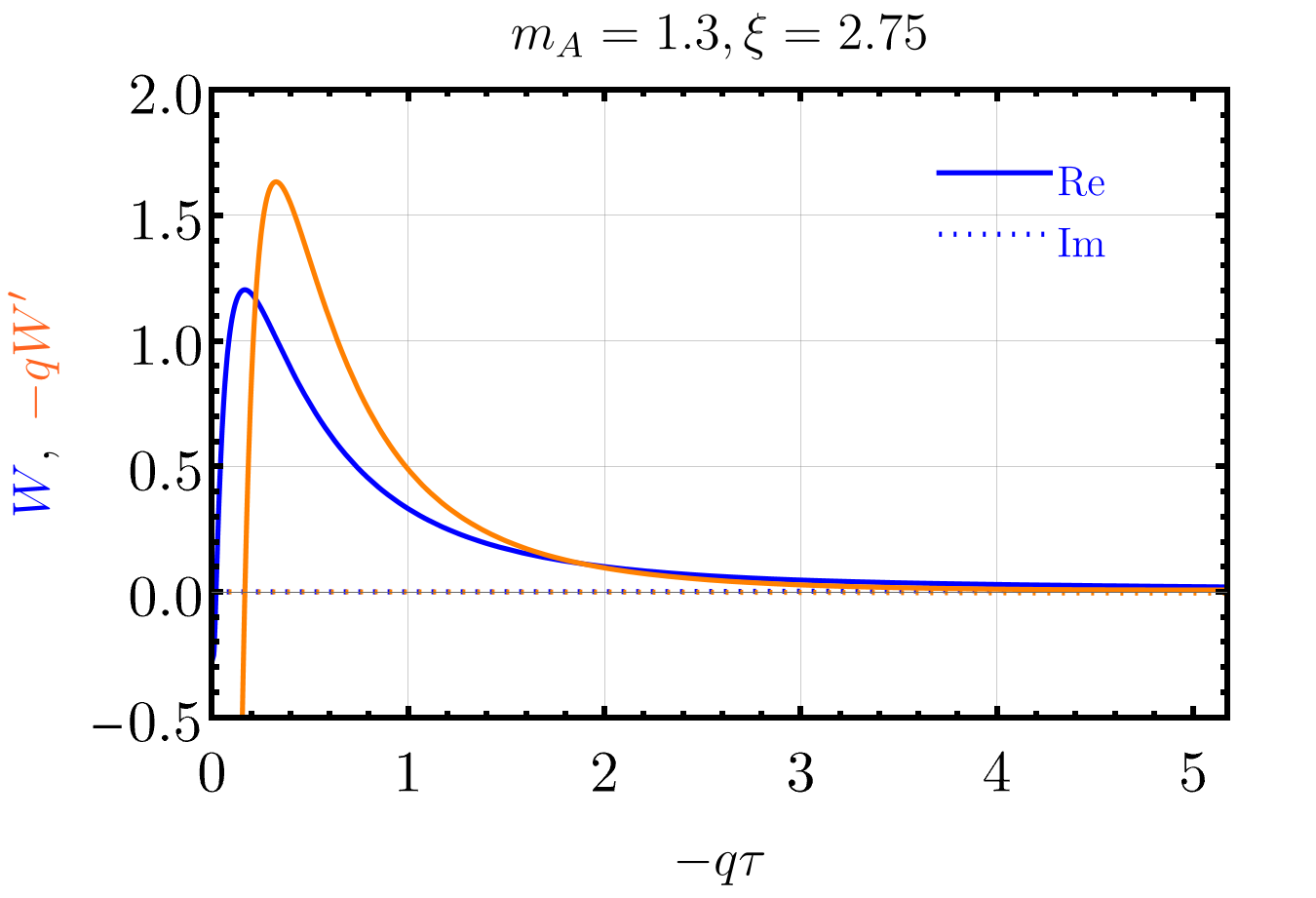}
\caption{The Whittaker function $W$ (blue) and its derivative $-q W'$ (orange). The functions are divided by the constant phase $e^{i\theta_0}$ of the 
Whittaker function evaluated at $ - q \tau = 0.4$. The real part of the two quantities shown in the solid curves 
are predominant in the time domain $ - q \tau \lesssim {\cal O}(1)$ over the imaginary ones (dotted curves).
}
\label{fig:WWp}
\end{center}
\end{figure}

The numerical analysis of the mode function demonstrates that the real mode function is a good approximation, and we can use
the simplified four-point function formalism to deduce the trispectrum. 
The real mode function approximation should be satisfied in the time region of the gauge boson mode $A_+$ dominantly produced.  
The energy density of gauge mode $A_+$ rises with time 
tells that the gauge boson production becomes effective when 
\begin{equation}  
   - q \tau \lesssim  \xi + \sqrt{ \xi^2 - \left( \frac{m_A} { H} \right)^2}  \, ,\label{hardcut-off}
\end{equation}  
which sets the lower limit of the time integration in eq.~(\ref{eq:sourceF1}) in the numerical study (details are given in \cite{Niu:2022quw}).
Furthermore, the integration peaks at $- q \tau \sim 0.1 - 1 $, as shown in \cref{fig:integrand}. In \cref{fig:integrand}, we choose $q=1$, and other
values of $q$ are tested, and the same conclusion holds.
This guides us to inspect the real mode function approximation in the regime of $- q \tau < {\cal O } (1)$ and 
find that the mode function $A_+$ and $A_+^\prime$ have 
a phase close to change, which can be rotated away. The real part is predominant and is about one or two orders of magnitude larger 
than the imaginary one. The examples for massless and massive 
gauge boson modes are shown in \cref{fig:WWp}. We obtain the phase of the Whittaker function 
$e^{i\theta_0} =  W_{\mp i\xi, i\mu}(2iq\tau )   / |  W_{\mp i\xi, i\mu}(2iq\tau )    |$
at $ - q \tau = 0.4$, strip out the constant phase of $W$ and $W'$, and find that in the time domain of $- q \tau \lesssim {\cal O}(1)$, 
the real part is dominant over the imaginary part. These plots show that the real mode function is a better approximation for the 
massless and massive cases.

Using the real mode function approximation to evaluate the correlation function, we should not include the imaginary part
of the mode functions, especially when we try to separate the parity even and odd contributions. 
The imaginary mode function would introduce an imaginary term in the time integration $\int d\tau \Im \delta \phi \, {\cal B}_2$.
It is a minor correction but contributes to the opposite parity signals. Even for the three-point function, it 
introduces fake parity odd signatures. For the four-point function, the amplitude of the parity odd signal usually is one or two orders of magnitude 
smaller than the parity even one. Then this fake parity odd contribution may contaminate the predictions. This contamination does not 
appear in the full in-in formalism due to $c.c$ in the eq.~(\ref{eq:4pointF1}). However, taking the real part of
$\int d\tau \Im \delta \phi {\cal B}_2$ with the complex mode function is a way to estimate the uncertainty of the real mode function approach by 
comparing the two results, rather than evaluating the time-consuming in-in formalism such as eq.~(\ref{eq:4pointF1}).
The parity even part is the dominant one, so that the approximation gives very precise results. Since the parity odd signals typically have cancellations 
among the different diagrams, the estimated uncertainties increase to $\sim 10\%$ from the numerical studies.

\section{Numerical results}
\label{sec:results}

In this section, we present the results of our analysis of the parity-even and odd parts of the curvature four-point function and show the 
shape function ${\cal T}({\bf k}_1,{\bf k}_2, {\bf k}_3, {\bf k}_4)$ in \cref{eq:4pointTff} as a function of the angle $\phi_r$ and model parameters $m_A/H$ and $\xi$. 

\subsection{Constraints from power spectrum and non-Gaussianity}

\begin{figure}[!ht] 
    \centering
      \includegraphics[width=0.6\textwidth]{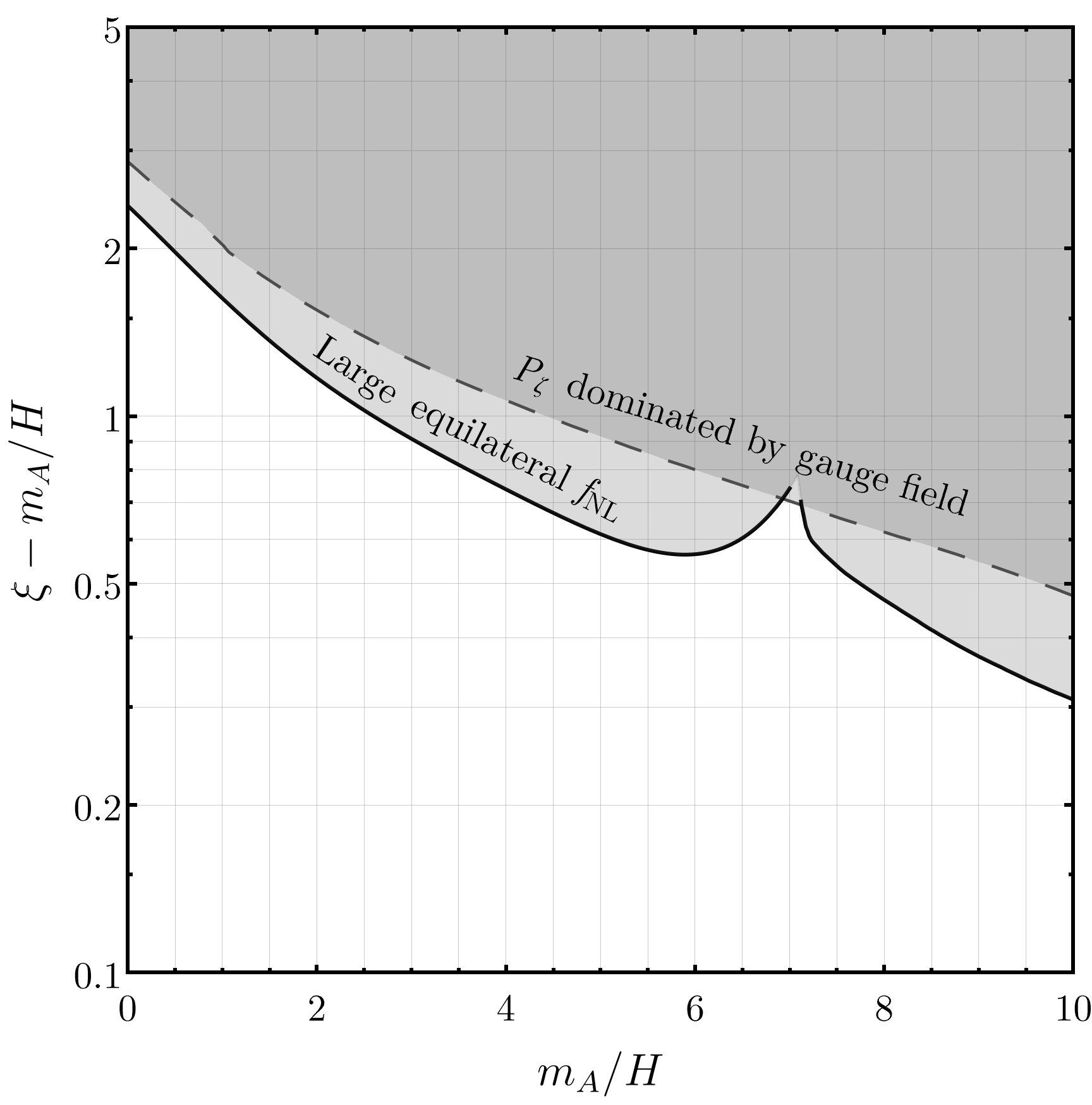}
    \caption{Constraints from scalar power spectrum and non-Gaussianity at the CMB scales on the parameter space of gauge boson production \cite{Niu:2022quw}. Shaded regions are excluded due to the large gauge boson one-loop radiative corrections to two- and three-point functions of curvature perturbations.
    }
         \label{fig:AllConstraints}
\end{figure}

We first review the constraints on the model parameters $m_A/H$ and $\xi$ from CMB observables related to the two- and three-point functions. These constraints are elaborated in Ref.~\cite{Niu:2022quw}. The scalar power spectrum is well-measured at the CMB scales by COBE normalization \cite{Bunn:1996py} and WMAP \cite{WMAP:2010qai}, $P_\zeta \simeq 2.5 \times 10^{-9}$. It is reasonable to expect that the power spectrum is dominated by vacuum modes at CMB scales. The parameter space where the one-loop correction in \cref{eq:zetazeta} dominates the tree-level contribution in \cref{eq:Pzeta} at the CMB scales is shown in 
\cref{fig:AllConstraints} with the dark shaded region labeled ``$P_{\zeta}$ dominated by gauge field''. 

The presence of the gauge field during inflation may induce sizable non-Gaussianity, which can be studied from the three-point function. So far, no evidence of non-Gaussianity has been observed at the CMB.
In equilateral configuration,  $|\mathbf{k_1}| = |\mathbf{k_2}| = |\mathbf{k_3}|$, the non-Gaussianity parameter $f_{\rm NL}$ is constrained by Planck 2018 data as  $f_{\rm NL}^{\rm eq} = -25 \pm 47$ at $68\%$ CL \cite{Planck:2019kim}. The parameter space violating this bound lies above the curves denoted with the label ``Large equilateral $f_{\rm NL}$'' in \cref{fig:AllConstraints}. The two segments of the boundary represent the limit from positive and negative bounds from $f_{\rm{NL}}^{\rm{eq}}$, respectively. 

The parameter space is also subject to bounds from strong backreaction from the gauge bosons and a large tensor-to-scalar ratio at the CMB scales. These bounds are weaker than the scalar non-Gaussianity bound, as found in Ref.~\cite{Niu:2022quw}, so it is not shown here. 

\subsection{Parity-odd and even trispectrum}

We now focus on the allowed parameter space in \cref{fig:AllConstraints} to investigate the parity-even and odd scalar four-point function. Based on our discussion in section~\ref{subsec:dominantreal}, we adopt the real mode function approximation in our numerical calculation of the four-point function. For the time integration, we set a hard cut-off following \cref{hardcut-off}. For the loop-momentum integration, we have verified that the integration converges for $q \sim O(1)-O(10)$. Therefore, we set a finite $O(10)$ cut-off. We use \texttt{Mathematica} for numerical calculations and parallelize the computations in high-performance cluster computers.
\begin{figure}[!ht]
    \centering
    \includegraphics[height=0.50\textwidth]{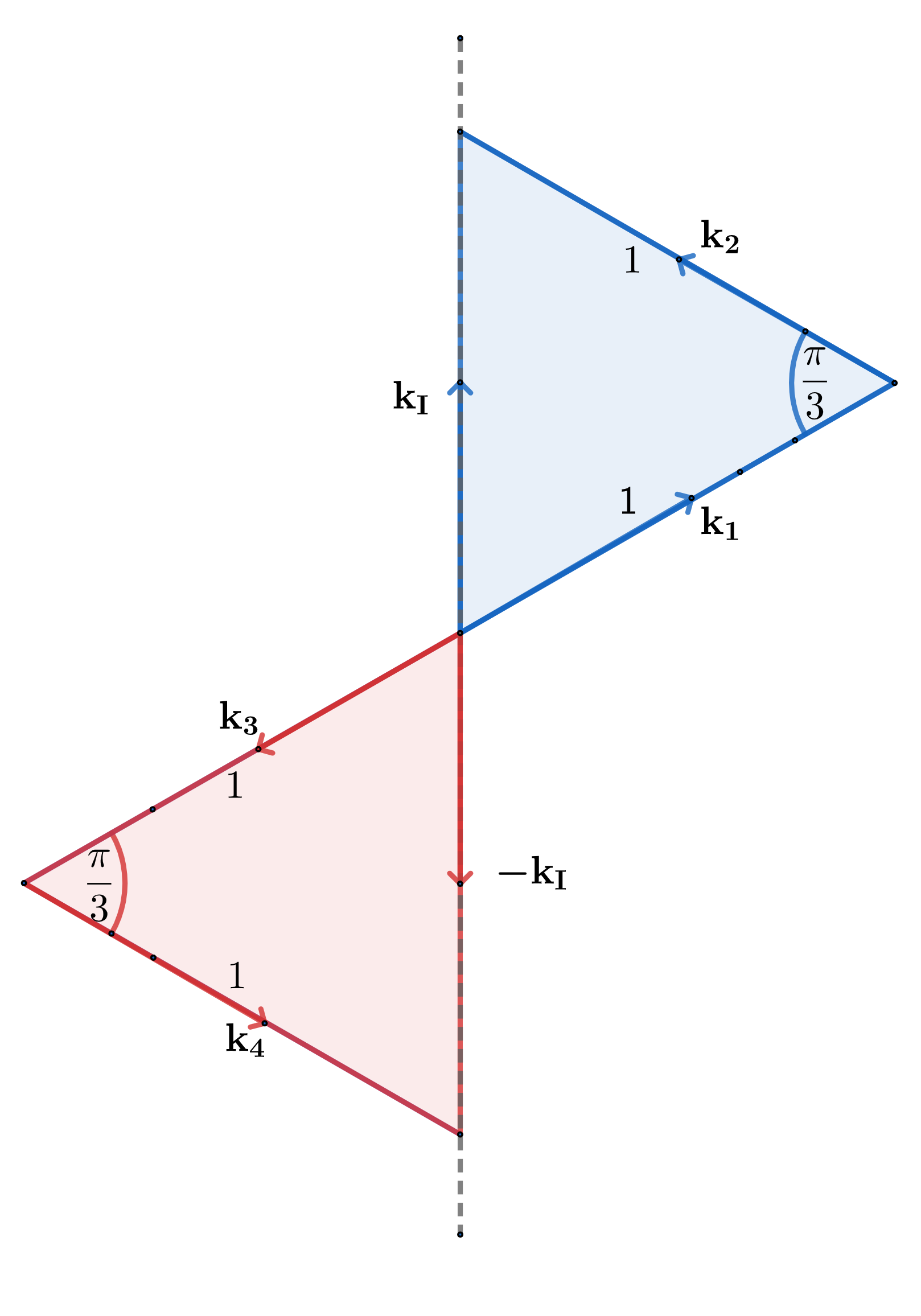}
    \hspace{0.0cm}
    \includegraphics[height=0.50\textwidth]{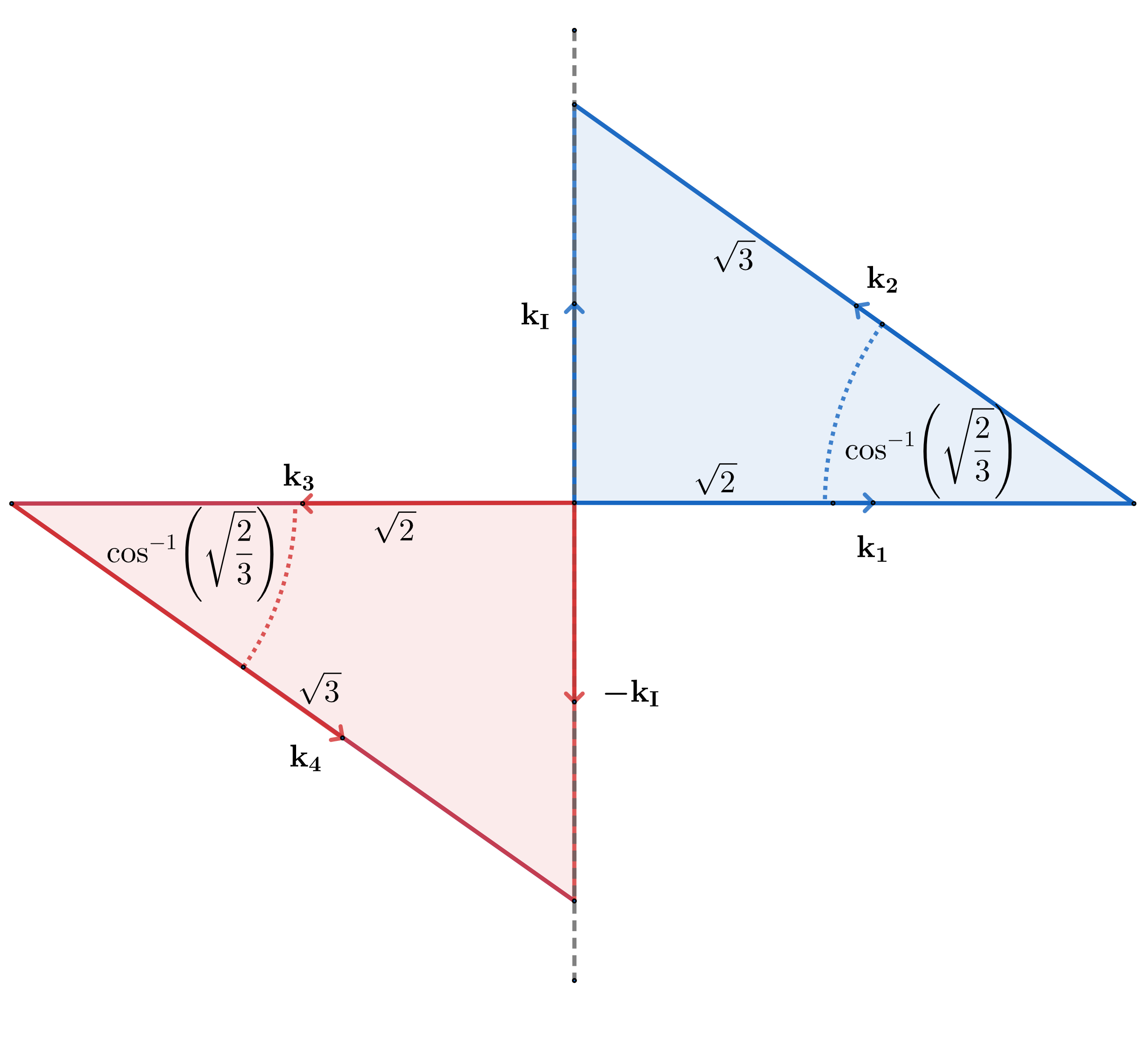}
    \caption{Momentum configurations I (left) and II (right). The red and the blue triangles are in different planes.}
    \label{mom.conf.}
\end{figure}

We specify two momentum configurations for our analysis,
\begin{itemize}
    \item \textbf{Config. I.} $|\mathbf{k_1}| = |\mathbf{k_2}| = |\mathbf{k_3}| = |\mathbf{k_4}| = 1$, $\theta_{12} = \theta_{34} = \dfrac{\pi}{3}$,
    \item \textbf{Config. II.} $\dfrac{|\mathbf{k_1}|}{\sqrt{2}} = \dfrac{|\mathbf{k_2}|}{\sqrt{3}}  = \dfrac{|\mathbf{k_3}|}{\sqrt{2}} = \dfrac{|\mathbf{k_4}|}{\sqrt{3}} = 1$, $\theta_{12} = \theta_{34} = \cos^{-1}\left(\sqrt{\dfrac{2}{3}}\right)\approx \dfrac{\pi}{5}$.
\end{itemize}
These configurations are illustrated in \cref{mom.conf.}. Note that the two triangles formed by $\mathbf{k_1}$, $\mathbf{k_2}$ and $\mathbf{k_3}$, $\mathbf{k_4}$ are in different planes, and the angle $\phi_r$ between the two planes is undetermined at this stage.

\subsubsection*{Massless case ($\boldsymbol{m_A =0}$ or $\boldsymbol{m_A \ll H}$)} 

\begin{figure}[!ht] 
    \centering
    \subfloat[]{        \includegraphics[width=0.33\textwidth]{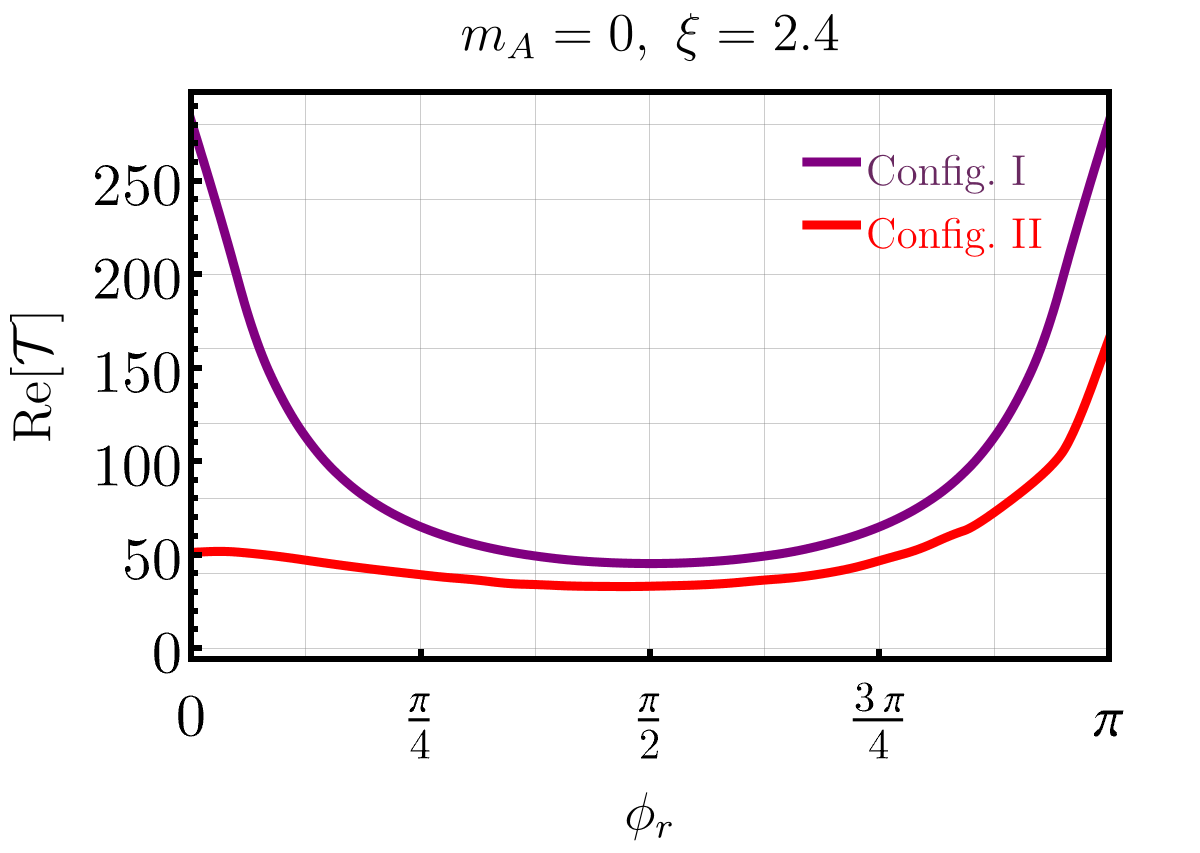}
    }\hspace*{-1.5em}
    \subfloat[]{        \includegraphics[width=0.335\textwidth]{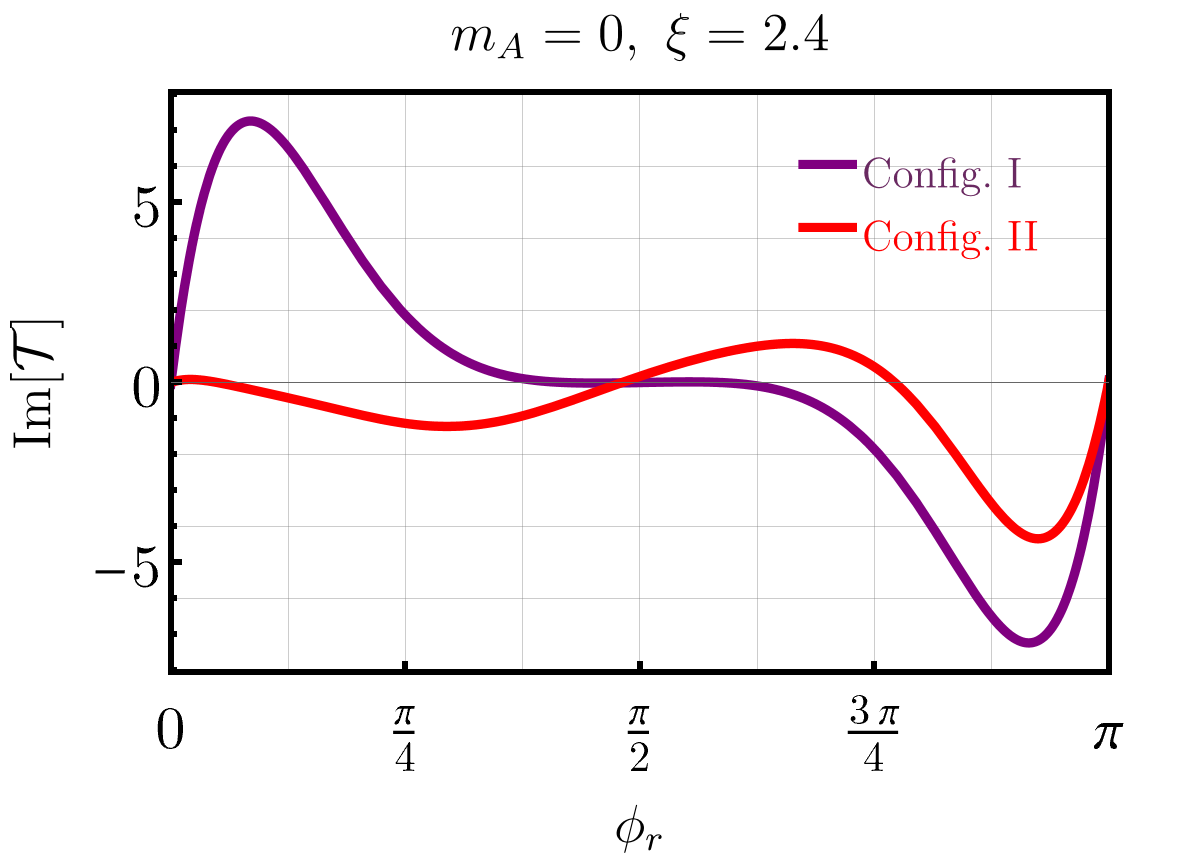}
    } \hspace*{-1.5em}
    \subfloat[\label{masslessratioTvsphi}]{        \includegraphics[width=0.36\textwidth]{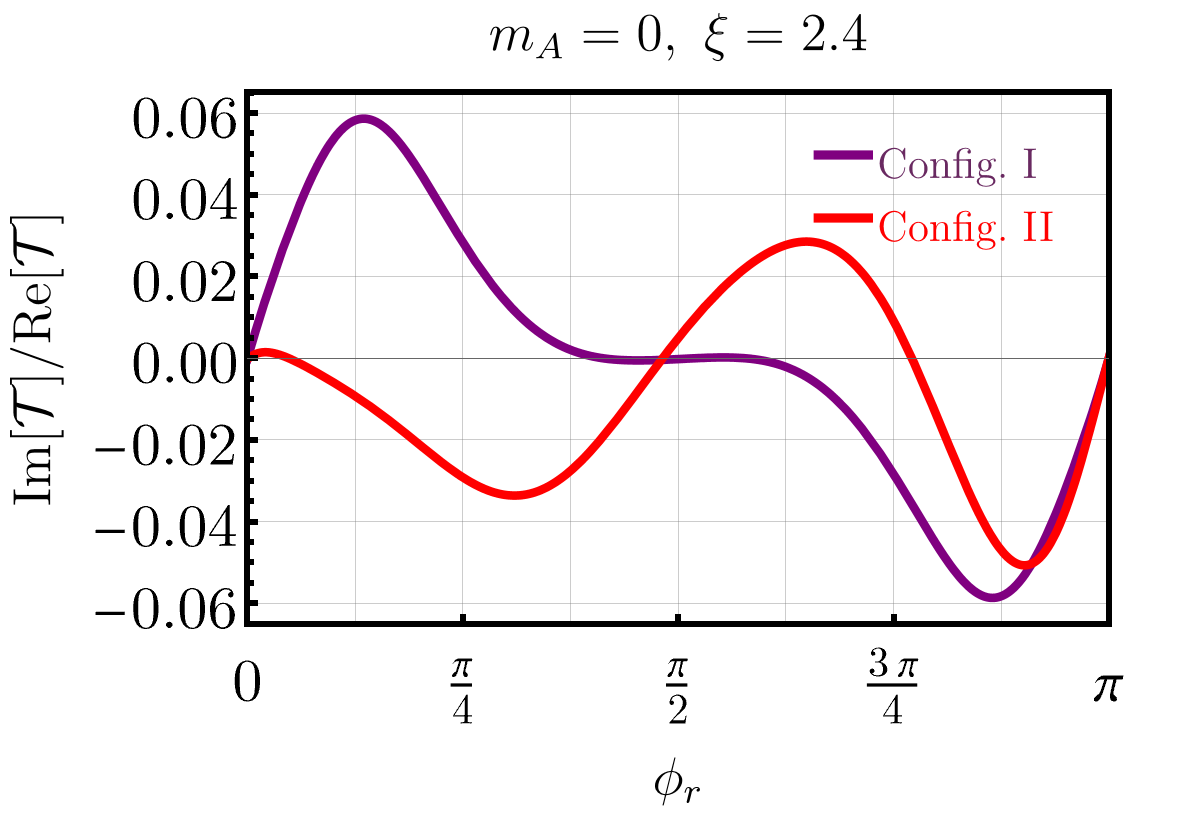}
    }
    \caption{Parity-even and odd signals and their ratio as a function of $\phi_r$ for massless vector boson. The purple curves correspond to Config.~I, and
    the red ones correspond to Config.~II. 
    } \label{phiscanmassless}
\end{figure}

We first investigate the parity-even and odd-signals in these two configurations as a function of the angle $\phi_r$ for the $m_A=0$ case. We choose a benchmark point $\xi=2.4$, corresponding to the largest allowed non-Gaussianity in \cref{fig:AllConstraints}. The results are presented in \cref{phiscanmassless}. The parity-even (Re[$\mathcal{T}$])  signal is at least $\mc{O}(10)$, and can be as large as $O(100)$ near $\phi_r \to 0$ (for Config.~I) and $\phi_r \to \pi$ (for both Config.~I and II). Config.~I yields a dominant signal by an $O(1)$ factor compared to Config.~II. 
The parity-odd (Im[$\mathcal{T}$]) signal vanishes for $\phi_r=0, \pi/2$ and $\pi$. 
It is typically suppressed compared with the parity-even signal, 
$|\text{Im}[\mc{T}]|/\text{Re}[\mc{T}] < 6\%$. 

We then turn to see the dependence of the trispectrum on the chemical potential $\xi$. We restrict our scan to the region $1.55 \leq \xi \leq 2.4$ where the uncertainty associated with using the dominant real mode function lies below $1\%$. We fix the angle $\phi_r$ to be $\pi/12$ and $3\pi/5$ for Config.~I and Config.~II, respectively. As expected, the signals exponentially depend on $\xi$. Consequently, $O(10) \sim O(100)$  parity-even signals occur for large $\xi$, which is close to the upper bound from $f_{\text{NL}}$ as shown in \cref{fig:AllConstraints}. Config.~I consistently yields a slightly larger signal than Config.~II. We observe similar behavior for the parity-odd signal. The ratio of the parity-odd to parity-even signal is up to  $\mc{O}(1)\%$ in the region of interest.
\begin{figure}[!ht] 
    \centering
    \subfloat[]{        \includegraphics[width=0.35\textwidth]{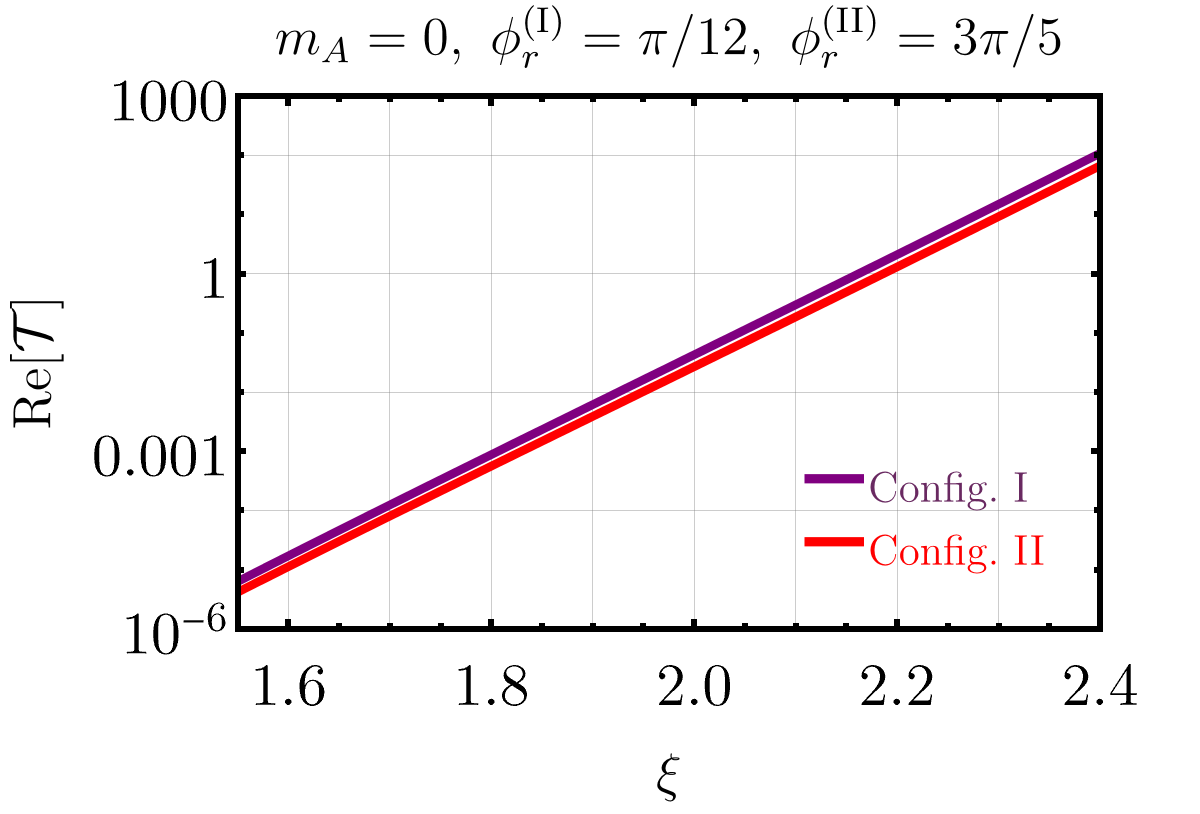}
    } \hspace*{-1.5em}
    \subfloat[]{        \includegraphics[width=0.35\textwidth]{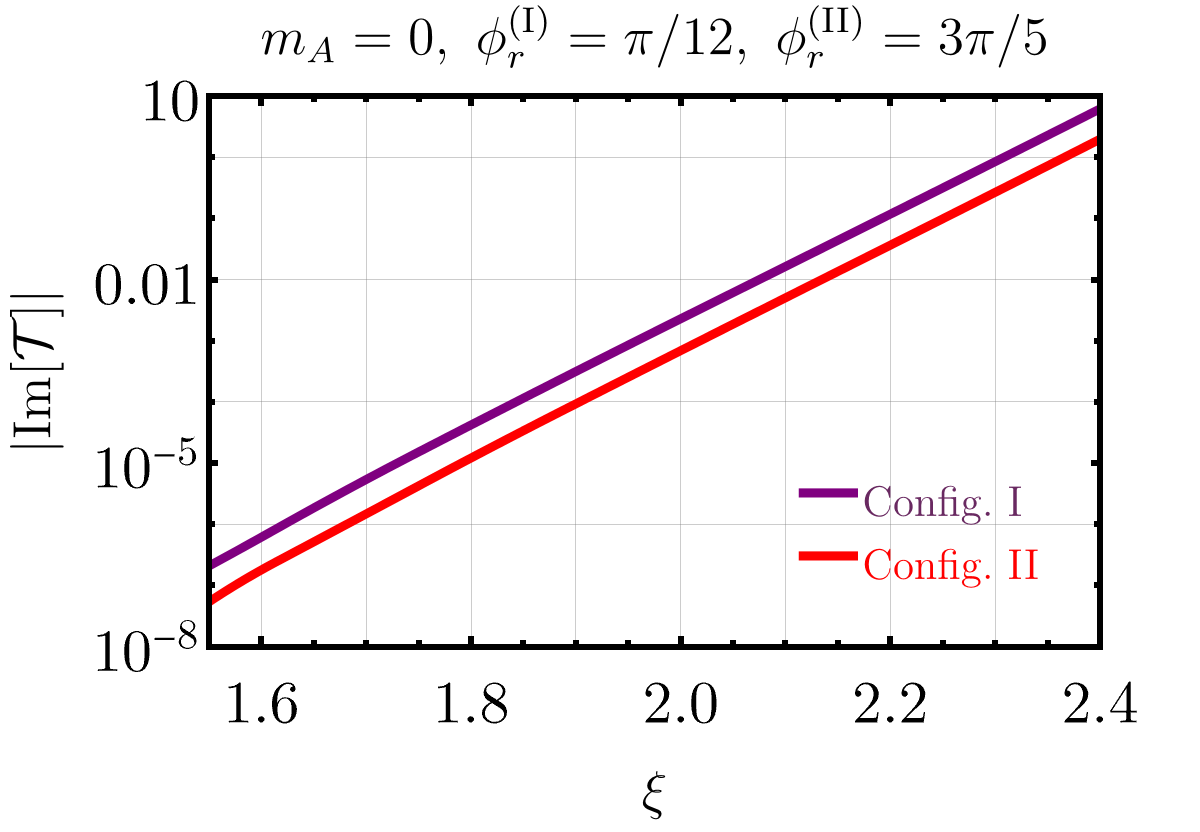}
    } \hspace*{-1.5em}
    \subfloat[\label{masslessratioTvsxi}]{        \includegraphics[width=0.36\textwidth]{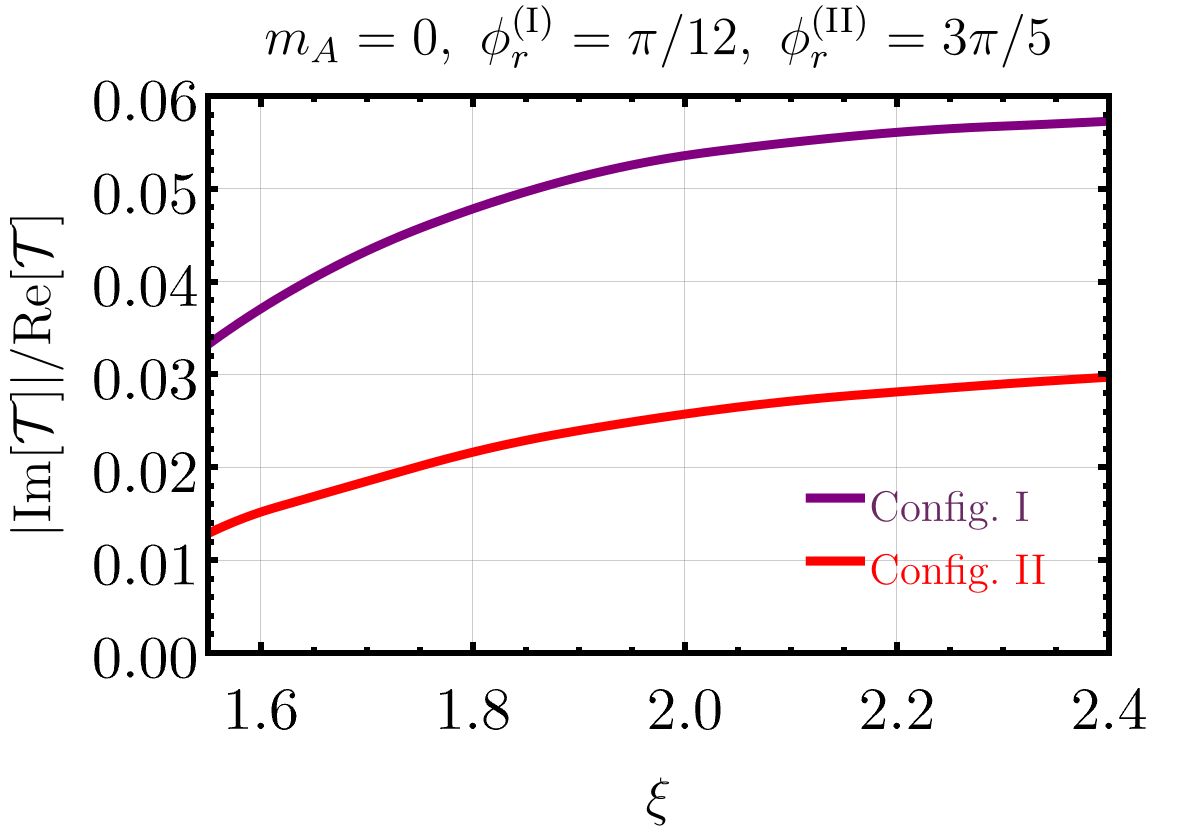}
    }
    \caption{Parity-even and odd signals as a function of $\xi$ for massless vector boson.} \label{xiscanmassless} 
\end{figure}

\subsubsection*{Massive case  ($\boldsymbol{m_A \sim H}$)}
\begin{figure}[!ht] 
    \centering
    \subfloat[]{        \includegraphics[width=0.335\textwidth]{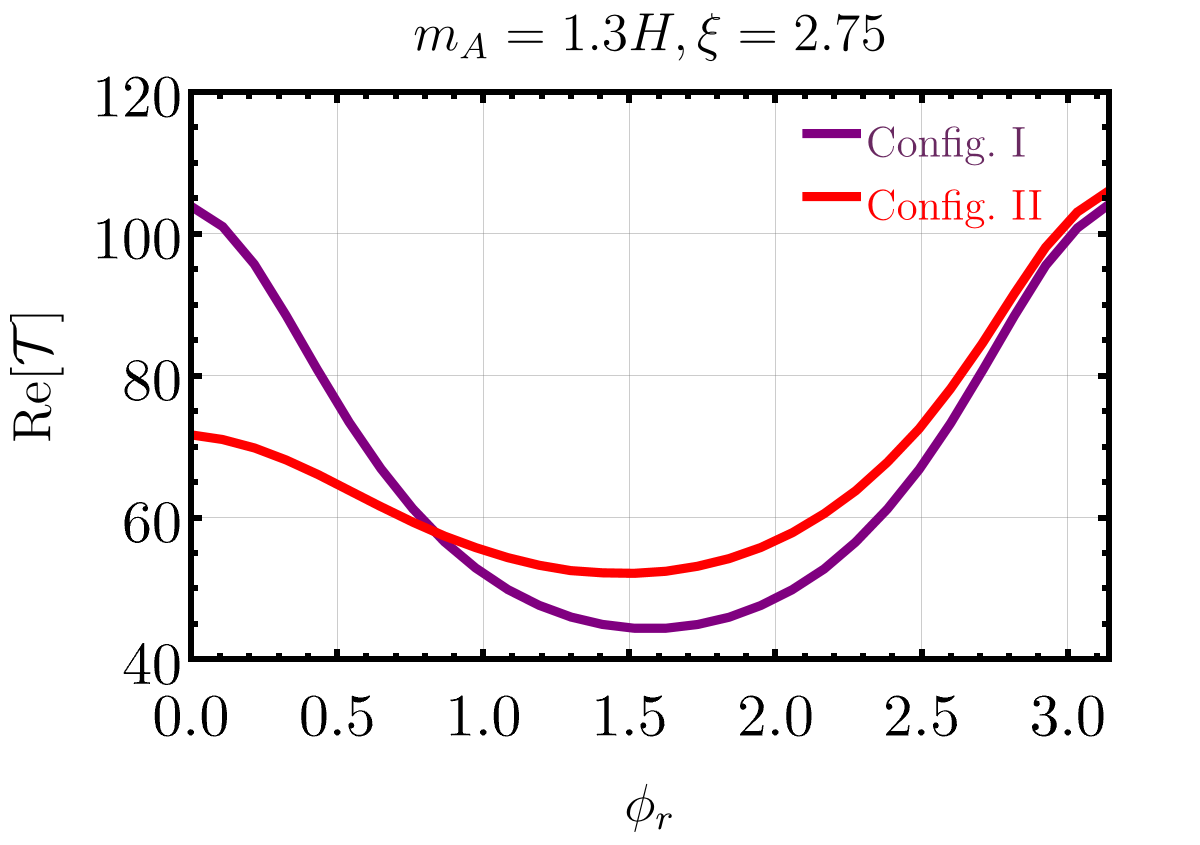}
    }\hspace*{-1.5em}
    \subfloat[]{        \includegraphics[width=0.34\textwidth]{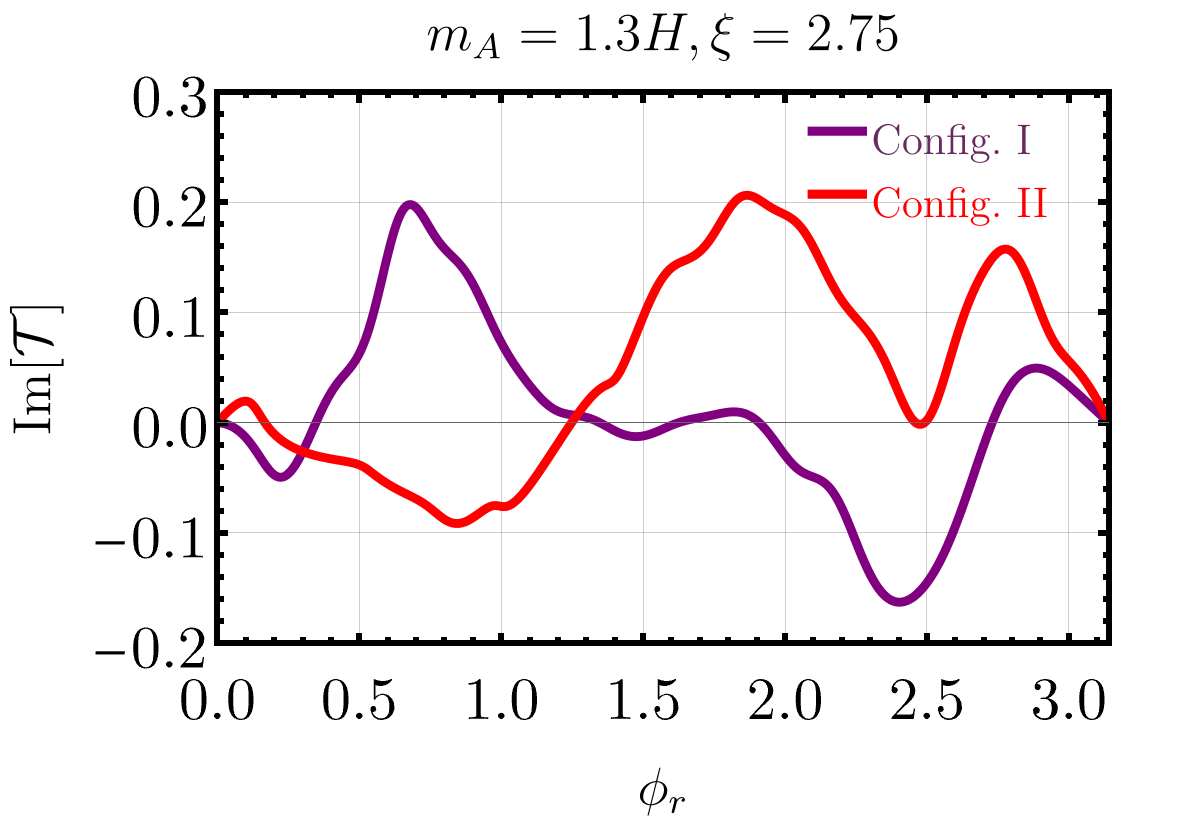}
    } \hspace*{-1.5em}
    \subfloat[\label{massiveratioTvsphi}]{        \includegraphics[width=0.37\textwidth]{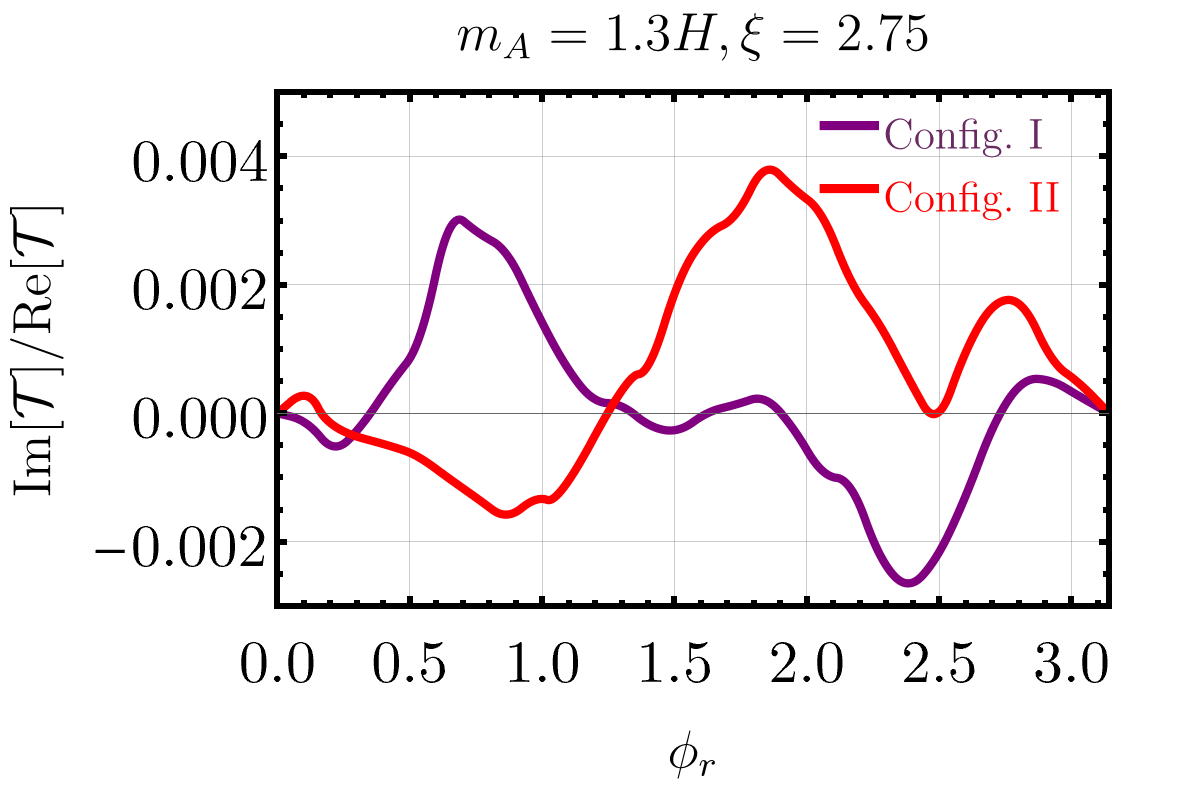}
    }
    \caption{Parity-even and odd signals and their ratio as a function of $\phi_r$ for massive vector boson.} \label{phiscanmassive}
\end{figure}

We choose a benchmark point $m_A=1.3H, \xi=2.75$, which produces the same $f_{\rm NL}$ as the benchmark point for the massless case. The parity-even and odd signals as a function of $\phi_r$ are shown in \cref{phiscanmassive}. While the parity-even signal can be $O(10) \sim O(100)$, the parity-odd signal is typically suppressed by three orders of magnitude. Compared with the massless case, the parity-even signal is comparable, but the parity-odd signal is weaker in the massive case.
Also, the curves for the parity odd signals are not as smooth as the ones in the massless case because of a larger uncertainty of the real mode approximation
for the massive case. 

\begin{figure}[!ht] 
    \centering
    \subfloat[]{        \includegraphics[width=0.35\textwidth]{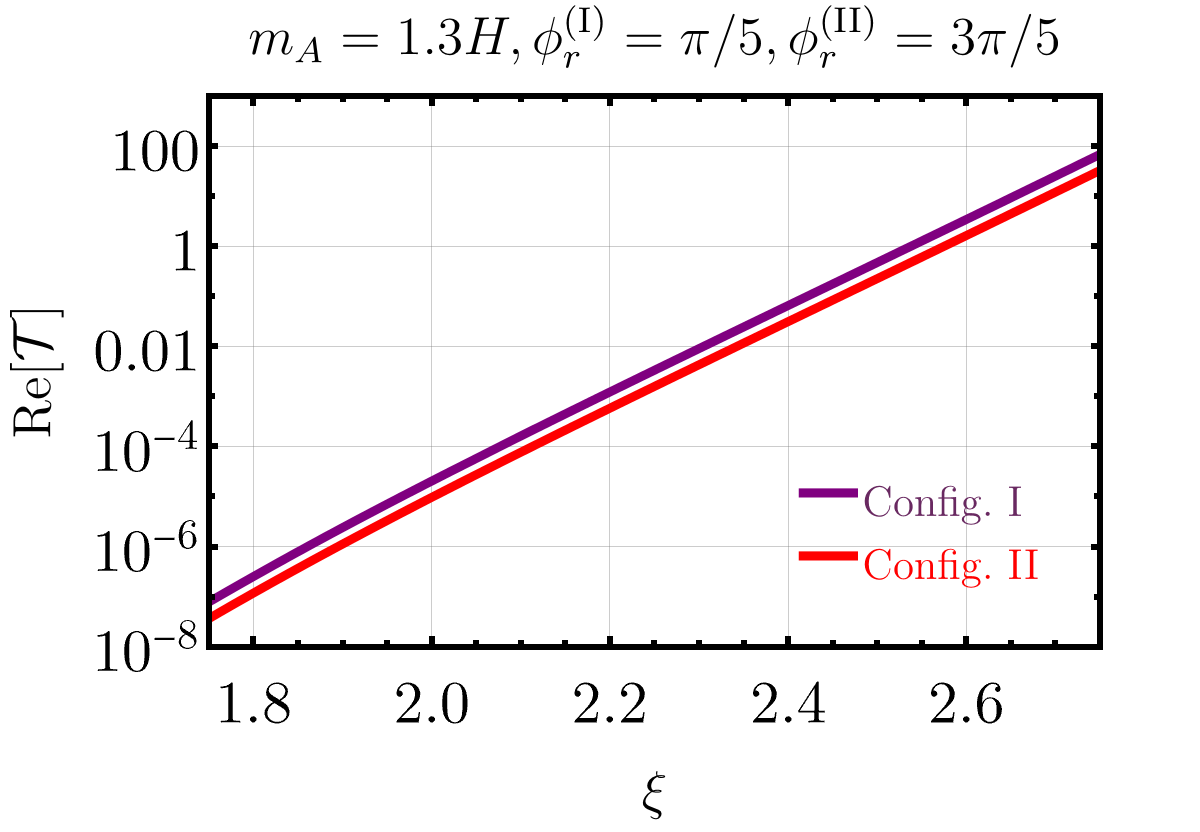}
    } \hspace*{-1.5em}
    \subfloat[]{        \includegraphics[width=0.35\textwidth]{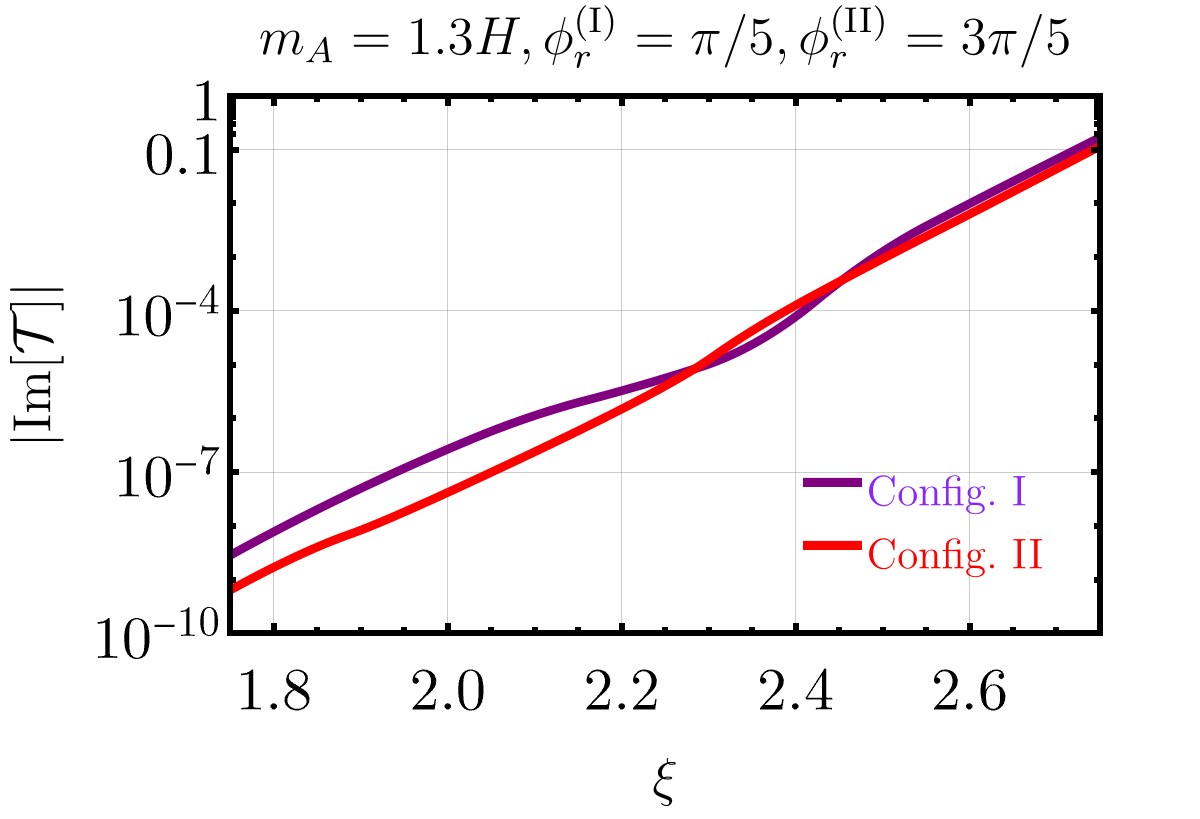}
    } \hspace*{-1.5em}
    \subfloat[\label{massiveratioTvsxi}]{        \includegraphics[width=0.36\textwidth]{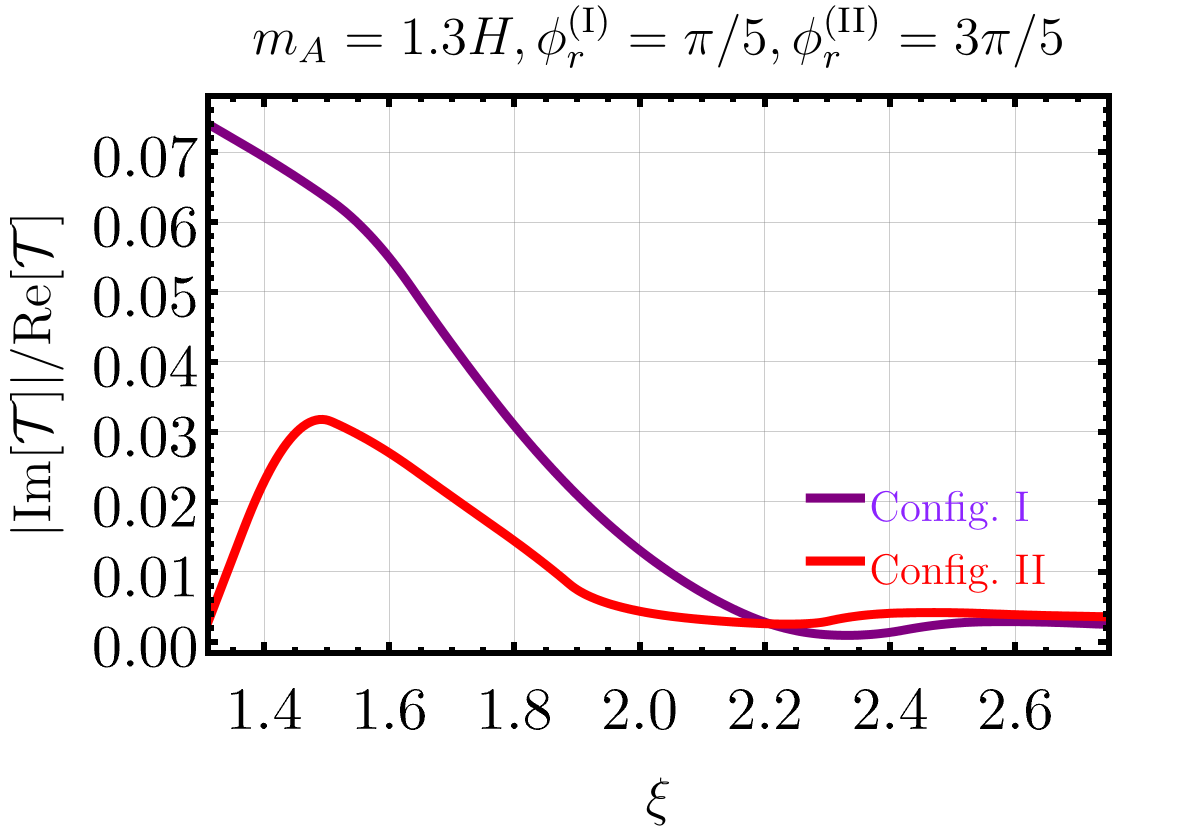}
    }
    \caption{Parity-even and odd signals as a function of $\xi$ for massive vector boson. 
    }
    \label{xiscanmassive}
\end{figure}

The dependence of the trispectrum on the chemical potential is illustrated in \cref{xiscanmassive}. We set $\phi_r = \pi/5$ and $\phi_r = 3\pi/5$ for Config.~I and Config.~II, respectively. 
The parity-even signals for the two momentum configurations are of the same order, with the Config.~I signal slightly dominating. The signal increases exponentially with $\xi$ and reaches $O(10)$ at large $\xi$. The parity-odd signal is $O(1\%)$ of its parity-even counterpart at smaller $\xi$ but becomes further suppressed as $\xi$ increases. In contrast, the ratio of odd to even parity signals consistently remained $O(1\%)$ in the massless case. This suggests that the massive gauge bosons produce a weaker parity-odd signal, even for benchmark points with similar values of the non-Gaussianity parameter $f_{\rm NL}^{\rm eq}$.

\begin{figure}[!ht] 
    \centering
    \subfloat[]{        \includegraphics[width=0.35\textwidth]{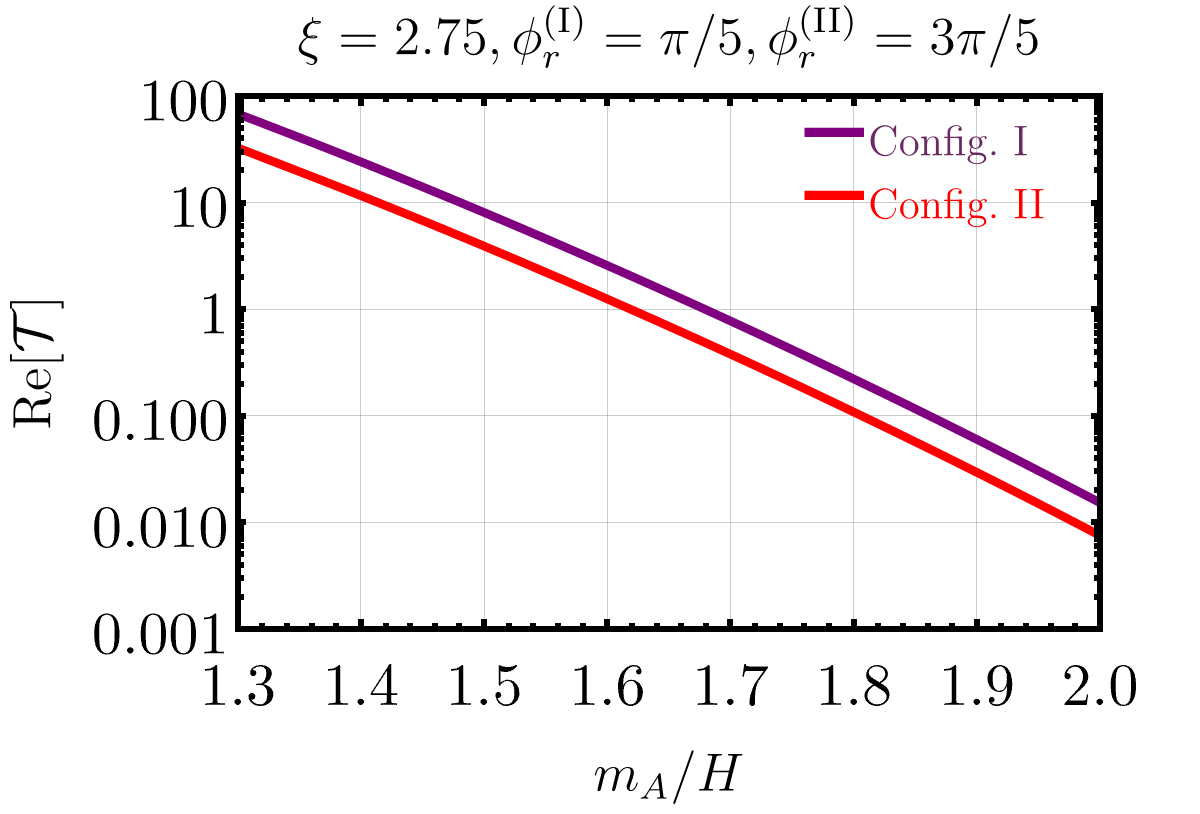}
    } \hspace*{-1.5em}
    \subfloat[]{        \includegraphics[width=0.35\textwidth]{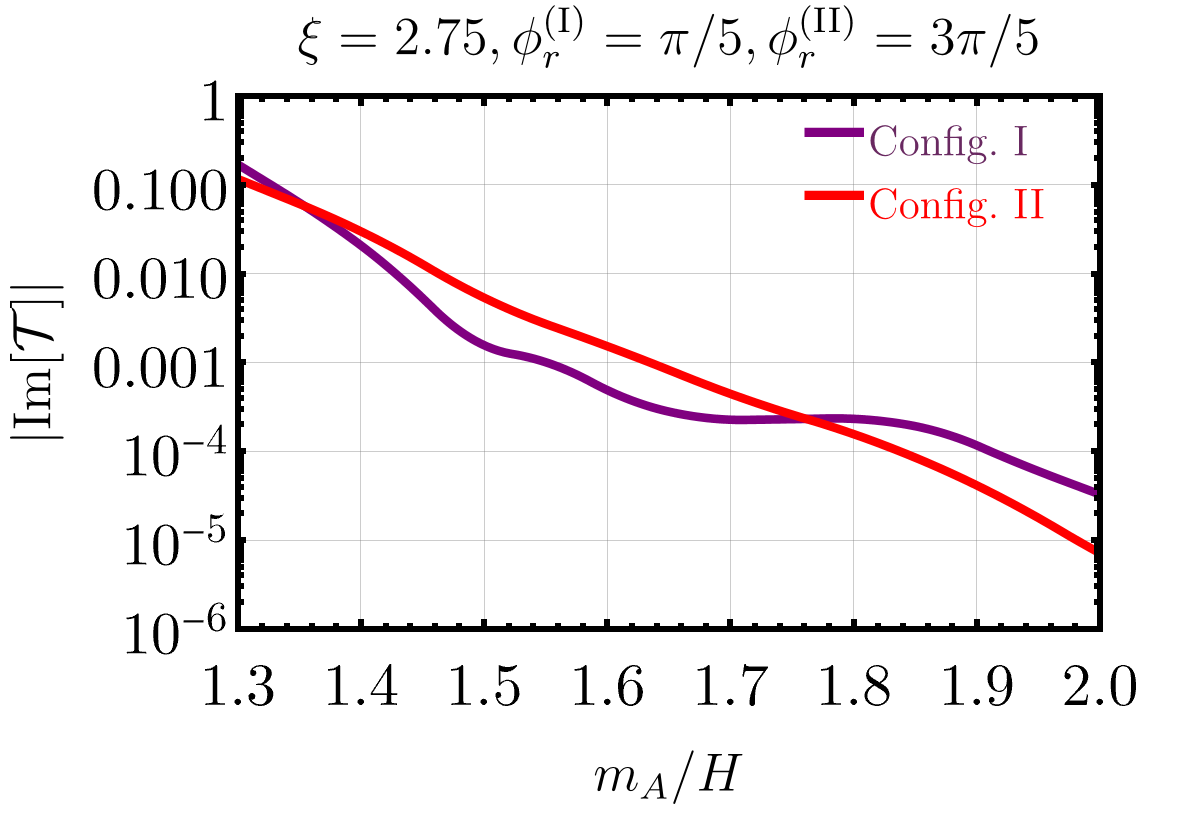}
    } \hspace*{-1.5em}
    \subfloat[\label{massiveratioTvsMass}]{        \includegraphics[width=0.36\textwidth]{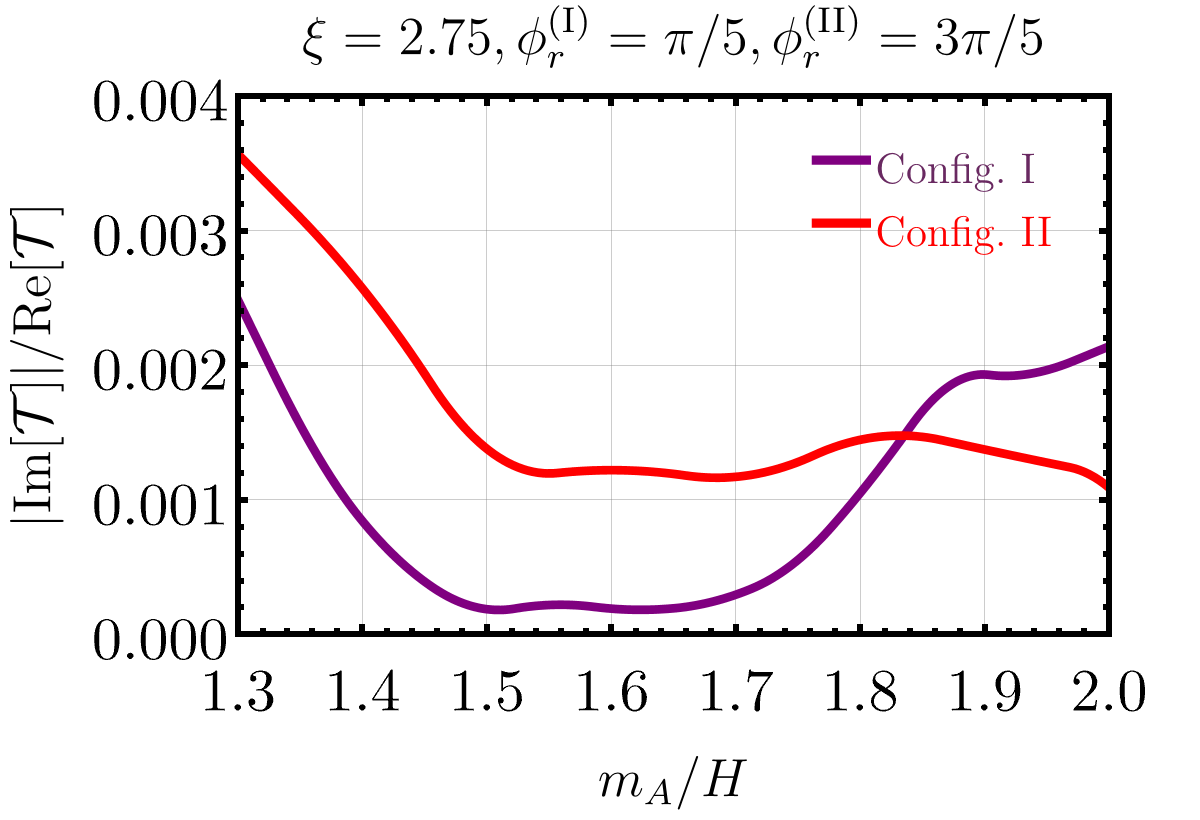}
    }
    \caption{Parity-even and odd signals as a function of $m_A/H$ for massive vector boson.} \label{massscanmassive}
\end{figure}

To further explore the mass dependence, we set $\xi$ at $2.75$ and increase $m_A/H$ from $1.3$ to $2$. The result is shown in \cref{massscanmassive}. The parity-odd and even signals drop with increasing mass, while their ratio is about $O(0.1\%)$ or weaker.

\section{Conclusion}\label{sec:conclusion}
We have investigated the parity-odd and even four-point functions in axion inflation, where a vector boson couples to the axion-like inflaton through a Chern-Simons coupling $\phi F \tilde{F}$. One of the transverse modes of massless and massive gauge bosons can be efficiently produced due to this coupling, which leaves an imprint on inflationary fluctuations. Unlike many other scenarios discussed in the literature, the contribution of the vector bosons to primordial perturbations can appear only at the loop level. Using the in-in formalism, we studied the one-loop diagrams for the four-point function. We have shown that the complicated in-in calculations can be vastly simplified by exploiting the fact that the complex gauge mode functions can be approximated as real ones
up to a global phase and that the real part of the mode function remains dominant after rephasing in the region where particle production effects are important. The real mode function approximation lends a powerful tool to evaluate functions numerically and can be applied to other relevant scenarios.

Since two- and three-point functions are inherently parity-conserving because of their coplanar topology, the four-point function provides the simplest configuration for studying parity violation. Its real and imaginary parts correspond to parity-even and parity-odd signals. 
The parity-odd signal would give a strong indication of parity violation in the early universe, as also contemplated by many BSM scenarios explaining the baryon asymmetry of the universe. We have shown that the parity-even trispectrum in axion inflation for both massless and massive gauge bosons can be as large as $O(10)$, in regions where the non-Gaussianity parameter $f_{\rm NL}^{\rm eq}$ from three-point statistics is also $O(10)$. The most stringent bound on the parity-even signal based on Planck 2018 data is Re[$\mathcal{T}$] $< 1700$ for given shapes \cite{Marzouk:2022utf}, hence the prediction from the axion inflation model can be probed in upcoming experiments. 

We have found that the parity-odd signal is non-vanishing and can be up to $O(1\%) - O(10\%)$ of its parity-even counterpart for massless gauge bosons. The signal is typically within $O(0.1\%) - O(10\%)$ of the parity-even signal for massive gauge bosons. It is an encouraging result, particularly in light of the recent hint of observing a parity-odd trispectrum in BOSS galaxy data \cite{Hou:2022wfj,Philcox:2022hkh}. It would be interesting to 
use the parity-odd four-point function of BOSS galaxies to probe the axion inflation model.

\acknowledgments
We are grateful to Jiamin Hou, Pierre Sikivie, and Zachary Slepian
for the helpful discussion.
This work was supported in part by the U.S. Department of Energy under grant DE-SC0022148 at the University of Florida. MHR acknowledges financial support from the STFC Consolidated Grant ST/T000775/1. The authors acknowledge the University of Florida Research Computing for providing computational resources and support that have contributed to the research results reported in this publication. We also acknowledge the use of the IRIDIS High-Performance Computing Facility and associated support services at the University of Southampton in the completion of this work.  

\newpage

\appendix

\section{4-point function}
\label{app:4point}

The second diagram in \cref{fig:Feynman} contributes to four-point function as 
\begin{eqnarray}
    && \langle  \zeta \left( \tau_0 , {\bf k}_1 \right)   \zeta ( \tau_0 , {\bf k}_2 ) 
         \zeta ( \tau_0 , {\bf k}_3 )  \zeta ( \tau_0 , {\bf k}_4 )  \rangle_{(F2)}^\prime
      \nonumber
      \\
       &=&  i^4    \left( - \frac{H}{\dot \phi_0 } \right)^4   \frac{H^4} { 4 (k_1 k_2 k_3 k_4)^{3/2}}  \frac{1}{\Lambda^4}
        \int_{-\infty}^0 d\tau_4  \int_{\infty}^{\tau_4} d\tau_3
        \int_{-\infty}^{\tau_3} d\tau_2  \int_{-\infty}^{\tau_2} d\tau_1 
       \int\frac{ d^3 q_1}{ ( 2\pi )^3} 
      \nonumber
      \\
      && 
          \boldsymbol{\epsilon}_+ ( {\bf q}_1 )  \cdot 
          \boldsymbol{\epsilon}_+ ( - {\bf q}_2  )   \, 
          \boldsymbol{\epsilon}_+ ( {\bf q}_2 )  \cdot 
          \boldsymbol{\epsilon}_+ ( - {\bf q}_3  )  \, 
          \boldsymbol{\epsilon}_+ ( {\bf q}_3 )  \cdot 
          \boldsymbol{\epsilon}_+ ( - {\bf q}_4  )  \,  
          \boldsymbol{\epsilon}_+ ( {\bf q}_4 )  \cdot 
          \boldsymbol{\epsilon}_+ ( - {\bf q}_1  )  \,  
      \nonumber
      \\
      &&
    \bigg[
      \delta\phi_{k_1}( \tau_1 )
      \delta\phi_{k_2}( \tau_2 )
      \delta\phi_{k_3}( \tau_3 )
     (  \delta\phi_{k_4}( \tau_4 )
      -   \delta\phi_{k_4}^*( \tau_4 ) )
      {\cal B}_2 ( q_1 , q_2, \tau_1 ) 
      {\cal B}_2 ( \underline{q_2 } , q_3 ,  \tau_2 ) 
      {\cal B}_2 (  \underline{q_3} , {q_4 } ,  \tau_3 ) 
      {\cal B}_2 (  \underline{q_4} , \underline{q_1 } ,  \tau_4 ) 
      \nonumber
      \\
      &&
      - \delta\phi_{k_1}( \tau_1 )
      \delta\phi_{k_2}( \tau_2 )
      \delta\phi_{k_3}^*( \tau_3 )
     (  \delta\phi_{k_4}( \tau_4 )
      -   \delta\phi_{k_4}^*( \tau_4 ) )
      {\cal B}_2 ( q_1 , q_2, \tau_1 ) 
      {\cal B}_2 ( \underline{q_2 } , q_3 ,  \tau_2 ) 
      {\cal B}_2 (  {q_4} , \underline{q_1}  ,  \tau_4 ) 
      {\cal B}_2 (  \underline{q_3} , \underline{q_4 } ,  \tau_3 ) 
      \nonumber
      \\
      &&
      - \delta\phi_{k_1}( \tau_1 )
      \delta\phi_{k_2}^*( \tau_2 )
      \delta\phi_{k_3}( \tau_3 )
     (  \delta\phi_{k_4}( \tau_4 )
      -   \delta\phi_{k_4}^*( \tau_4 ) )
      {\cal B}_2 ( q_1 , q_2, \tau_1 ) 
      {\cal B}_2 (  {q_3} , {q_4 } ,  \tau_3 ) 
      {\cal B}_2 (  \underline{q_4} , \underline{q_1}  ,  \tau_4 ) 
      {\cal B}_2 ( \underline{q_2 } , \underline{q_3} ,  \tau_2 ) 
      \nonumber
      \\
      &&
      + \delta\phi_{k_1}( \tau_1 )
      \delta\phi_{k_2}^*( \tau_2 )
      \delta\phi_{k_3}^*( \tau_3 )
     (  \delta\phi_{k_4}( \tau_4 )
      -   \delta\phi_{k_4}^*( \tau_4 ) )
      {\cal B}_2 ( q_1 , q_2, \tau_1 ) 
      {\cal B}_2 (  {q_4} , \underline{q_1}  ,  \tau_4 ) 
      {\cal B}_2 (  {q_3} , \underline{q_4 } ,  \tau_3 ) 
      {\cal B}_2 ( \underline{q_2 } , \underline{q_3} ,  \tau_2 ) 
      \nonumber
      \\
      &&
      +  {\rm 23 \, perms }
      + \rm{c.c.}
       \bigg] \, .
\label{eq:4pointF2}
\end{eqnarray}
The four momentum has the relation as  
\begin{equation}
   {\bf q}_2  = {\bf q}_1  - {\bf k}_1 \, , 
      \quad
   {\bf q}_3  = {\bf q}_1  - {\bf k}_1 - {\bf k}_2 \, ,
      \quad
   {\bf q}_4  = {\bf q}_1  + {\bf k}_1 \, .
   \label{eq:q234F2}
\end{equation}

\noindent The third diagram in \cref{fig:Feynman} contribute to four-point function as 
\begin{eqnarray}
    && \langle  \zeta \left( \tau_0 , {\bf k}_1 \right)   \zeta ( \tau_0 , {\bf k}_2 ) 
         \zeta ( \tau_0 , {\bf k}_3 )  \zeta ( \tau_0 , {\bf k}_4 )  \rangle_{(F3)}^\prime
      \nonumber
      \\
       &=&  i^4    \left( - \frac{H}{\dot \phi_0 } \right)^4   \frac{H^4} { 4 (k_1 k_2 k_3 k_4)^{3/2}}  \frac{1}{\Lambda^4}
        \int_{-\infty}^0 d\tau_4  \int_{\infty}^{\tau_4} d\tau_3
        \int_{-\infty}^{\tau_3} d\tau_2  \int_{-\infty}^{\tau_2} d\tau_1 
       \int\frac{ d^3 q_1}{ ( 2\pi )^3} 
      \nonumber
      \\
      && 
          \boldsymbol{\epsilon}_+ ( {\bf q}_1 )  \cdot 
          \boldsymbol{\epsilon}_+ ( - {\bf q}_3  )   \, 
          \boldsymbol{\epsilon}_+ ( {\bf q}_3 )  \cdot 
          \boldsymbol{\epsilon}_+ ( - {\bf q}_2  )  \, 
          \boldsymbol{\epsilon}_+ ( {\bf q}_2 )  \cdot 
          \boldsymbol{\epsilon}_+ ( - {\bf q}_4  )  \,  
          \boldsymbol{\epsilon}_+ ( {\bf q}_4 )  \cdot 
          \boldsymbol{\epsilon}_+ ( - {\bf q}_1  )  \,  
      \nonumber
      \\
      &&
    \bigg[
      \delta\phi_{k_1}( \tau_1 )
      \delta\phi_{k_2}( \tau_2 )
      \delta\phi_{k_3}( \tau_3 )
     (  \delta\phi_{k_4}( \tau_4 )
      -   \delta\phi_{k_4}^*( \tau_4 ) )
      {\cal B}_2 ( q_1 , q_3, \tau_1 ) 
      {\cal B}_2 ( {q_2 } , q_4 ,  \tau_2 ) 
      {\cal B}_2 (  \underline{q_3} , \underline{q_2 } ,  \tau_3 ) 
      {\cal B}_2 (  \underline{q_4} , \underline{q_1 } ,  \tau_4 ) 
      \nonumber
      \\
      &&
      - \delta\phi_{k_1}( \tau_1 )
      \delta\phi_{k_2}( \tau_2 )
      \delta\phi_{k_3}^*( \tau_3 )
     (  \delta\phi_{k_4}( \tau_4 )
      -   \delta\phi_{k_4}^*( \tau_4 ) )
      {\cal B}_2 ( q_1 , q_3, \tau_1 ) 
      {\cal B}_2 ( {q_2 } , q_4 ,  \tau_2 ) 
      {\cal B}_2 (  \underline{q_4} , \underline{q_1}  ,  \tau_4 ) 
      {\cal B}_2 (  \underline{q_3} , \underline{q_2 } ,  \tau_3 ) 
      \nonumber
      \\
      &&
      - \delta\phi_{k_1}( \tau_1 )
      \delta\phi_{k_2}^*( \tau_2 )
      \delta\phi_{k_3}( \tau_3 )
     (  \delta\phi_{k_4}( \tau_4 )
      -   \delta\phi_{k_4}^*( \tau_4 ) )
      {\cal B}_2 ( q_1 , q_3, \tau_1 ) 
      {\cal B}_2 (  \underline{q_3} , {q_2 } ,  \tau_3 ) 
      {\cal B}_2 (  {q_4} , \underline{q_1}  ,  \tau_4 ) 
      {\cal B}_2 ( \underline{q_2 } , \underline{q_4} ,  \tau_2 ) 
      \nonumber
      \\
      &&
      + \delta\phi_{k_1}( \tau_1 )
      \delta\phi_{k_2}^*( \tau_2 )
      \delta\phi_{k_3}^*( \tau_3 )
     (  \delta\phi_{k_4}( \tau_4 )
      -   \delta\phi_{k_4}^*( \tau_4 ) )
      {\cal B}_2 ( q_1 , q_3, \tau_1 ) 
      {\cal B}_2 (  {q_4} , \underline{q_1}  ,  \tau_4 ) 
      {\cal B}_2 (  \underline{q_3} , {q_2 } ,  \tau_3 ) 
      {\cal B}_2 ( \underline{q_2 } , \underline{q_4} ,  \tau_2 ) 
      \nonumber
      \\
      &&
      +  {\rm 23 \, perms }
      + \rm{c.c.}
       \bigg] \, .
\label{eq:4pointF3}
\end{eqnarray}

\noindent The four momentum has the relation as  
\begin{equation}
   {\bf q}_2  = {\bf q}_1  - {\bf k}_1 - {\bf k}_2 \, ,
      \quad
   {\bf q}_3  = {\bf q}_1  - {\bf k}_1 \, , 
      \quad
   {\bf q}_4  = {\bf q}_1  + {\bf k}_1 \, .
   \label{eq:q234F3}
\end{equation}

\noindent When considering the real mode $A_+$ approximation, the four-point function simplifies to 
\begin{eqnarray}
    && \langle  \zeta ( \tau_0 , {\bf k}_1 )   \zeta ( \tau_0 , {\bf k}_2 ) 
         \zeta ( \tau_0 , {\bf k}_3 )  \zeta ( \tau_0 , {\bf k}_4 )  \rangle_{(F2)}^\prime
       =     \left(  \frac{H}{\dot \phi_0 } \right)^4   \frac{2^4 H^4} { 4 (k_1 k_2 k_3 k_4)^{3/2}}  \frac{1}{\Lambda^4}
      \nonumber
      \\
      && \times 
       \int\frac{ d^3 q_1}{ ( 2\pi )^3} 
          \boldsymbol{\epsilon}_+ ( {\bf q}_1 )  \cdot 
          \boldsymbol{\epsilon}_+ ( - {\bf q}_3  )   \, 
          \boldsymbol{\epsilon}_+ ( {\bf q}_3 )  \cdot 
          \boldsymbol{\epsilon}_+ ( - {\bf q}_2  )  \, 
          \boldsymbol{\epsilon}_+ ( {\bf q}_2 )  \cdot 
          \boldsymbol{\epsilon}_+ ( - {\bf q}_4  )  \,  
          \boldsymbol{\epsilon}_+ ( {\bf q}_4 )  \cdot 
          \boldsymbol{\epsilon}_+ ( - {\bf q}_1  )  \,  
      \nonumber
      \\
      &&
      \times 
       \int_{-\infty}^{0} d\tau_1 \Im \delta\phi_{k_1}( \tau_1 ) {\cal B}_2 ( q_1 , q_2, \tau_1 )  \, 
       \int_{-\infty}^{0} d\tau_2 \Im \delta\phi_{k_2}( \tau_2 ) {\cal B}_2 ( \underline{q_2} , q_3, \tau_2 ) 
      \nonumber
      \\
      &&
      \times 
       \int_{-\infty}^{0} d\tau_3 \Im \delta\phi_{k_3}( \tau_3 ) {\cal B}_2 ( \underline{q_3} , {q_4}, \tau_3 ) 
       \int_{-\infty}^{0} d\tau_4 \Im \delta\phi_{k_4}( \tau_4 ) {\cal B}_2 ( \underline{q_4} , \underline{q_1}, \tau_4 ) 
      \, .
   \label{eq:sourceF2}
\end{eqnarray}

\begin{eqnarray}
    && \langle  \zeta ( \tau_0 , {\bf k}_1 )   \zeta ( \tau_0 , {\bf k}_2 ) 
         \zeta ( \tau_0 , {\bf k}_3 )  \zeta ( \tau_0 , {\bf k}_4 )  \rangle_{(F3)}^\prime
       =     \left(  \frac{H}{\dot \phi_0 } \right)^4   \frac{2^4 H^4} { 4 (k_1 k_2 k_3 k_4)^{3/2}}  \frac{1}{\Lambda^4}
      \nonumber
      \\
      && \times 
       \int\frac{ d^3 q_1}{ ( 2\pi )^3} 
          \boldsymbol{\epsilon}_+ ( {\bf q}_1 )  \cdot 
          \boldsymbol{\epsilon}_+ ( - {\bf q}_3  )   \, 
          \boldsymbol{\epsilon}_+ ( {\bf q}_3 )  \cdot 
          \boldsymbol{\epsilon}_+ ( - {\bf q}_2  )  \, 
          \boldsymbol{\epsilon}_+ ( {\bf q}_2 )  \cdot 
          \boldsymbol{\epsilon}_+ ( - {\bf q}_4  )  \,  
          \boldsymbol{\epsilon}_+ ( {\bf q}_4 )  \cdot 
          \boldsymbol{\epsilon}_+ ( - {\bf q}_1  )  \,  
      \nonumber
      \\
      &&
      \times 
       \int_{-\infty}^{0} d\tau_1 \Im \delta\phi_{k_1}( \tau_1 ) {\cal B}_2 ( q_1 , q_3, \tau_1 )  \, 
       \int_{-\infty}^{0} d\tau_2 \Im \delta\phi_{k_2}( \tau_2 ) {\cal B}_2 ( {q_2} , q_4, \tau_2 ) 
      \nonumber
      \\
      &&
      \times 
       \int_{-\infty}^{0} d\tau_3 \Im \delta\phi_{k_3}( \tau_3 ) {\cal B}_2 ( \underline{q_3} , \underline{q_2}, \tau_3 ) 
       \int_{-\infty}^{0} d\tau_4 \Im \delta\phi_{k_4}( \tau_4 ) {\cal B}_2 ( \underline{q_4} , \underline{q_1}, \tau_4 ) 
      \, .
   \label{eq:sourceF3}
\end{eqnarray}

\bibliography{references}

\providecommand{\href}[2]{#2}\begingroup\raggedright\begin{thebibliography}{100}

\bibitem{Brout:1977ix}
R.~Brout, F.~Englert and E.~Gunzig, \emph{{The Creation of the Universe as a
  Quantum Phenomenon}},
  \href{https://doi.org/10.1016/0003-4916(78)90176-8}{\emph{Annals Phys.}
  {\bfseries 115} (1978) 78}.

\bibitem{Starobinsky:1980te}
A.A.~Starobinsky, \emph{{A New Type of Isotropic Cosmological Models Without
  Singularity}},
  \href{https://doi.org/10.1016/0370-2693(80)90670-X}{\emph{Phys. Lett. B}
  {\bfseries 91} (1980) 99}.

\bibitem{Kazanas:1980tx}
D.~Kazanas, \emph{{Dynamics of the Universe and Spontaneous Symmetry
  Breaking}}, \href{https://doi.org/10.1086/183361}{\emph{Astrophys. J. Lett.}
  {\bfseries 241} (1980) L59}.

\bibitem{Sato:1980yn}
K.~Sato, \emph{{First Order Phase Transition of a Vacuum and Expansion of the
  Universe}}, {\emph{Mon. Not. Roy. Astron. Soc.} {\bfseries 195} (1981) 467}.

\bibitem{Guth:1980zm}
A.H.~Guth, \emph{{The Inflationary Universe: A Possible Solution to the Horizon
  and Flatness Problems}},
  \href{https://doi.org/10.1103/PhysRevD.23.347}{\emph{Phys. Rev. D} {\bfseries
  23} (1981) 347}.

\bibitem{Linde:1981mu}
A.D.~Linde, \emph{{A New Inflationary Universe Scenario: A Possible Solution of
  the Horizon, Flatness, Homogeneity, Isotropy and Primordial Monopole
  Problems}}, \href{https://doi.org/10.1016/0370-2693(82)91219-9}{\emph{Phys.
  Lett. B} {\bfseries 108} (1982) 389}.

\bibitem{Albrecht:1982wi}
A.~Albrecht and P.J.~Steinhardt, \emph{{Cosmology for Grand Unified Theories
  with Radiatively Induced Symmetry Breaking}},
  \href{https://doi.org/10.1103/PhysRevLett.48.1220}{\emph{Phys. Rev. Lett.}
  {\bfseries 48} (1982) 1220}.

\bibitem{Linde:1983gd}
A.D.~Linde, \emph{{Chaotic Inflation}},
  \href{https://doi.org/10.1016/0370-2693(83)90837-7}{\emph{Phys. Lett. B}
  {\bfseries 129} (1983) 177}.

\bibitem{Chen:2006nt}
X.~Chen, M.-x.~Huang, S.~Kachru and G.~Shiu, \emph{{Observational signatures
  and non-Gaussianities of general single field inflation}},
  \href{https://doi.org/10.1088/1475-7516/2007/01/002}{\emph{JCAP} {\bfseries
  01} (2007) 002} [\href{https://arxiv.org/abs/hep-th/0605045}{{\ttfamily
  hep-th/0605045}}].

\bibitem{Cheung:2007st}
C.~Cheung, P.~Creminelli, A.L.~Fitzpatrick, J.~Kaplan and L.~Senatore,
  \emph{{The Effective Field Theory of Inflation}},
  \href{https://doi.org/10.1088/1126-6708/2008/03/014}{\emph{JHEP} {\bfseries
  03} (2008) 014} [\href{https://arxiv.org/abs/0709.0293}{{\ttfamily
  0709.0293}}].

\bibitem{Adshead:2011jq}
P.~Adshead, C.~Dvorkin, W.~Hu and E.A.~Lim, \emph{{Non-Gaussianity from Step
  Features in the Inflationary Potential}},
  \href{https://doi.org/10.1103/PhysRevD.85.023531}{\emph{Phys. Rev. D}
  {\bfseries 85} (2012) 023531}
  [\href{https://arxiv.org/abs/1110.3050}{{\ttfamily 1110.3050}}].

\bibitem{Holman:2007na}
R.~Holman and A.J.~Tolley, \emph{{Enhanced Non-Gaussianity from Excited Initial
  States}}, \href{https://doi.org/10.1088/1475-7516/2008/05/001}{\emph{JCAP}
  {\bfseries 05} (2008) 001} [\href{https://arxiv.org/abs/0710.1302}{{\ttfamily
  0710.1302}}].

\bibitem{Meerburg:2009ys}
P.D.~Meerburg, J.P.~van~der Schaar and P.S.~Corasaniti, \emph{{Signatures of
  Initial State Modifications on Bispectrum Statistics}},
  \href{https://doi.org/10.1088/1475-7516/2009/05/018}{\emph{JCAP} {\bfseries
  05} (2009) 018} [\href{https://arxiv.org/abs/0901.4044}{{\ttfamily
  0901.4044}}].

\bibitem{Ashoorioon:2010xg}
A.~Ashoorioon and G.~Shiu, \emph{{A Note on Calm Excited States of Inflation}},
  \href{https://doi.org/10.1088/1475-7516/2011/03/025}{\emph{JCAP} {\bfseries
  03} (2011) 025} [\href{https://arxiv.org/abs/1012.3392}{{\ttfamily
  1012.3392}}].

\bibitem{Chen:2009zp}
X.~Chen and Y.~Wang, \emph{{Quasi-Single Field Inflation and
  Non-Gaussianities}},
  \href{https://doi.org/10.1088/1475-7516/2010/04/027}{\emph{JCAP} {\bfseries
  04} (2010) 027} [\href{https://arxiv.org/abs/0911.3380}{{\ttfamily
  0911.3380}}].

\bibitem{Chen:2009we}
X.~Chen and Y.~Wang, \emph{{Large non-Gaussianities with Intermediate Shapes
  from Quasi-Single Field Inflation}},
  \href{https://doi.org/10.1103/PhysRevD.81.063511}{\emph{Phys. Rev. D}
  {\bfseries 81} (2010) 063511}
  [\href{https://arxiv.org/abs/0909.0496}{{\ttfamily 0909.0496}}].

\bibitem{Baumann:2011nk}
D.~Baumann and D.~Green, \emph{{Signatures of Supersymmetry from the Early
  Universe}}, \href{https://doi.org/10.1103/PhysRevD.85.103520}{\emph{Phys.
  Rev. D} {\bfseries 85} (2012) 103520}
  [\href{https://arxiv.org/abs/1109.0292}{{\ttfamily 1109.0292}}].

\bibitem{Arkani-Hamed:2015bza}
N.~Arkani-Hamed and J.~Maldacena, \emph{{Cosmological Collider Physics}},
  \href{https://arxiv.org/abs/1503.08043}{{\ttfamily 1503.08043}}.

\bibitem{Chen:2016nrs}
X.~Chen, Y.~Wang and Z.-Z.~Xianyu, \emph{{Loop Corrections to Standard Model
  Fields in Inflation}},
  \href{https://doi.org/10.1007/JHEP08(2016)051}{\emph{JHEP} {\bfseries 08}
  (2016) 051} [\href{https://arxiv.org/abs/1604.07841}{{\ttfamily
  1604.07841}}].

\bibitem{Lee:2016vti}
H.~Lee, D.~Baumann and G.L.~Pimentel, \emph{{Non-Gaussianity as a Particle
  Detector}}, \href{https://doi.org/10.1007/JHEP12(2016)040}{\emph{JHEP}
  {\bfseries 12} (2016) 040}
  [\href{https://arxiv.org/abs/1607.03735}{{\ttfamily 1607.03735}}].

\bibitem{Meerburg:2016zdz}
P.D.~Meerburg, M.~M\"unchmeyer, J.B.~Mu\~noz and X.~Chen, \emph{{Prospects for
  Cosmological Collider Physics}},
  \href{https://doi.org/10.1088/1475-7516/2017/03/050}{\emph{JCAP} {\bfseries
  03} (2017) 050} [\href{https://arxiv.org/abs/1610.06559}{{\ttfamily
  1610.06559}}].

\bibitem{Chen:2016uwp}
X.~Chen, Y.~Wang and Z.-Z.~Xianyu, \emph{{Standard Model Background of the
  Cosmological Collider}},
  \href{https://doi.org/10.1103/PhysRevLett.118.261302}{\emph{Phys. Rev. Lett.}
  {\bfseries 118} (2017) 261302}
  [\href{https://arxiv.org/abs/1610.06597}{{\ttfamily 1610.06597}}].

\bibitem{Chen:2016hrz}
X.~Chen, Y.~Wang and Z.-Z.~Xianyu, \emph{{Standard Model Mass Spectrum in
  Inflationary Universe}},
  \href{https://doi.org/10.1007/JHEP04(2017)058}{\emph{JHEP} {\bfseries 04}
  (2017) 058} [\href{https://arxiv.org/abs/1612.08122}{{\ttfamily
  1612.08122}}].

\bibitem{An:2017hlx}
H.~An, M.~McAneny, A.K.~Ridgway and M.B.~Wise, \emph{{Quasi Single Field
  Inflation in the non-perturbative regime}},
  \href{https://doi.org/10.1007/JHEP06(2018)105}{\emph{JHEP} {\bfseries 06}
  (2018) 105} [\href{https://arxiv.org/abs/1706.09971}{{\ttfamily
  1706.09971}}].

\bibitem{Kumar:2017ecc}
S.~Kumar and R.~Sundrum, \emph{{Heavy-Lifting of Gauge Theories By Cosmic
  Inflation}}, \href{https://doi.org/10.1007/JHEP05(2018)011}{\emph{JHEP}
  {\bfseries 05} (2018) 011}
  [\href{https://arxiv.org/abs/1711.03988}{{\ttfamily 1711.03988}}].

\bibitem{Chen:2018xck}
X.~Chen, Y.~Wang and Z.-Z.~Xianyu, \emph{{Neutrino Signatures in Primordial
  Non-Gaussianities}},
  \href{https://doi.org/10.1007/JHEP09(2018)022}{\emph{JHEP} {\bfseries 09}
  (2018) 022} [\href{https://arxiv.org/abs/1805.02656}{{\ttfamily
  1805.02656}}].

\bibitem{Wu:2018lmx}
Y.-P.~Wu, \emph{{Higgs as heavy-lifted physics during inflation}},
  \href{https://doi.org/10.1007/JHEP04(2019)125}{\emph{JHEP} {\bfseries 04}
  (2019) 125} [\href{https://arxiv.org/abs/1812.10654}{{\ttfamily
  1812.10654}}].

\bibitem{Li:2019ves}
L.~Li, T.~Nakama, C.M.~Sou, Y.~Wang and S.~Zhou, \emph{{Gravitational
  Production of Superheavy Dark Matter and Associated Cosmological
  Signatures}}, \href{https://doi.org/10.1007/JHEP07(2019)067}{\emph{JHEP}
  {\bfseries 07} (2019) 067}
  [\href{https://arxiv.org/abs/1903.08842}{{\ttfamily 1903.08842}}].

\bibitem{Lu:2019tjj}
S.~Lu, Y.~Wang and Z.-Z.~Xianyu, \emph{{A Cosmological Higgs Collider}},
  \href{https://doi.org/10.1007/JHEP02(2020)011}{\emph{JHEP} {\bfseries 02}
  (2020) 011} [\href{https://arxiv.org/abs/1907.07390}{{\ttfamily
  1907.07390}}].

\bibitem{Hook:2019zxa}
A.~Hook, J.~Huang and D.~Racco, \emph{{Searches for other vacua. Part II. A new
  Higgstory at the cosmological collider}},
  \href{https://doi.org/10.1007/JHEP01(2020)105}{\emph{JHEP} {\bfseries 01}
  (2020) 105} [\href{https://arxiv.org/abs/1907.10624}{{\ttfamily
  1907.10624}}].

\bibitem{Hook:2019vcn}
A.~Hook, J.~Huang and D.~Racco, \emph{{Minimal signatures of the Standard Model
  in non-Gaussianities}},
  \href{https://doi.org/10.1103/PhysRevD.101.023519}{\emph{Phys. Rev. D}
  {\bfseries 101} (2020) 023519}
  [\href{https://arxiv.org/abs/1908.00019}{{\ttfamily 1908.00019}}].

\bibitem{Kumar:2019ebj}
S.~Kumar and R.~Sundrum, \emph{{Cosmological Collider Physics and the
  Curvaton}}, \href{https://doi.org/10.1007/JHEP04(2020)077}{\emph{JHEP}
  {\bfseries 04} (2020) 077}
  [\href{https://arxiv.org/abs/1908.11378}{{\ttfamily 1908.11378}}].

\bibitem{Wang:2019gbi}
L.-T.~Wang and Z.-Z.~Xianyu, \emph{{In Search of Large Signals at the
  Cosmological Collider}},
  \href{https://doi.org/10.1007/JHEP02(2020)044}{\emph{JHEP} {\bfseries 02}
  (2020) 044} [\href{https://arxiv.org/abs/1910.12876}{{\ttfamily
  1910.12876}}].

\bibitem{Wang:2020uic}
Y.~Wang and Y.~Zhu, \emph{{Cosmological Collider Signatures of Massive Vectors
  from Non-Gaussian Gravitational Waves}},
  \href{https://doi.org/10.1088/1475-7516/2020/04/049}{\emph{JCAP} {\bfseries
  04} (2020) 049} [\href{https://arxiv.org/abs/2001.03879}{{\ttfamily
  2001.03879}}].

\bibitem{Wang:2020ioa}
L.-T.~Wang and Z.-Z.~Xianyu, \emph{{Gauge Boson Signals at the Cosmological
  Collider}}, \href{https://doi.org/10.1007/JHEP11(2020)082}{\emph{JHEP}
  {\bfseries 11} (2020) 082}
  [\href{https://arxiv.org/abs/2004.02887}{{\ttfamily 2004.02887}}].

\bibitem{Maru:2021ezc}
N.~Maru and A.~Okawa, \emph{{Non-Gaussianity from $X, Y$ gauge bosons in
  Cosmological Collider Physics}},
  \href{https://arxiv.org/abs/2101.10634}{{\ttfamily 2101.10634}}.

\bibitem{Lu:2021gso}
S.~Lu, \emph{{Axion isocurvature collider}},
  \href{https://doi.org/10.1007/JHEP04(2022)157}{\emph{JHEP} {\bfseries 04}
  (2022) 157} [\href{https://arxiv.org/abs/2103.05958}{{\ttfamily
  2103.05958}}].

\bibitem{Wang:2021qez}
L.-T.~Wang, Z.-Z.~Xianyu and Y.-M.~Zhong, \emph{{Precision calculation of
  inflation correlators at one loop}},
  \href{https://doi.org/10.1007/JHEP02(2022)085}{\emph{JHEP} {\bfseries 02}
  (2022) 085} [\href{https://arxiv.org/abs/2109.14635}{{\ttfamily
  2109.14635}}].

\bibitem{Tong:2021wai}
X.~Tong, Y.~Wang and Y.~Zhu, \emph{{Cutting rule for cosmological collider
  signals: a bulk evolution perspective}},
  \href{https://doi.org/10.1007/JHEP03(2022)181}{\emph{JHEP} {\bfseries 03}
  (2022) 181} [\href{https://arxiv.org/abs/2112.03448}{{\ttfamily
  2112.03448}}].

\bibitem{Cui:2021iie}
Y.~Cui and Z.-Z.~Xianyu, \emph{{Probing Leptogenesis with the Cosmological
  Collider}}, \href{https://doi.org/10.1103/PhysRevLett.129.111301}{\emph{Phys.
  Rev. Lett.} {\bfseries 129} (2022) 111301}
  [\href{https://arxiv.org/abs/2112.10793}{{\ttfamily 2112.10793}}].

\bibitem{Pinol:2021aun}
L.~Pinol, S.~Aoki, S.~Renaux-Petel and M.~Yamaguchi, \emph{{Inflationary flavor
  oscillations and the cosmic spectroscopy}},
  \href{https://arxiv.org/abs/2112.05710}{{\ttfamily 2112.05710}}.

\bibitem{Tong:2022cdz}
X.~Tong and Z.-Z.~Xianyu, \emph{{Large spin-2 signals at the cosmological
  collider}}, \href{https://doi.org/10.1007/JHEP10(2022)194}{\emph{JHEP}
  {\bfseries 10} (2022) 194}
  [\href{https://arxiv.org/abs/2203.06349}{{\ttfamily 2203.06349}}].

\bibitem{Reece:2022soh}
M.~Reece, L.-T.~Wang and Z.-Z.~Xianyu, \emph{{Large-Field Inflation and the
  Cosmological Collider}},  \href{https://arxiv.org/abs/2204.11869}{{\ttfamily
  2204.11869}}.

\bibitem{Jazayeri:2022kjy}
S.~Jazayeri and S.~Renaux-Petel, \emph{{Cosmological Bootstrap in Slow
  Motion}},  \href{https://arxiv.org/abs/2205.10340}{{\ttfamily 2205.10340}}.

\bibitem{Pimentel:2022fsc}
G.L.~Pimentel and D.-G.~Wang, \emph{{Boostless cosmological collider
  bootstrap}}, \href{https://doi.org/10.1007/JHEP10(2022)177}{\emph{JHEP}
  {\bfseries 10} (2022) 177}
  [\href{https://arxiv.org/abs/2205.00013}{{\ttfamily 2205.00013}}].

\bibitem{Chen:2022vzh}
X.~Chen, R.~Ebadi and S.~Kumar, \emph{{Classical cosmological collider physics
  and primordial features}},
  \href{https://doi.org/10.1088/1475-7516/2022/08/083}{\emph{JCAP} {\bfseries
  08} (2022) 083} [\href{https://arxiv.org/abs/2205.01107}{{\ttfamily
  2205.01107}}].

\bibitem{Qin:2022lva}
Z.~Qin and Z.-Z.~Xianyu, \emph{{Phase information in cosmological collider
  signals}}, \href{https://doi.org/10.1007/JHEP10(2022)192}{\emph{JHEP}
  {\bfseries 10} (2022) 192}
  [\href{https://arxiv.org/abs/2205.01692}{{\ttfamily 2205.01692}}].

\bibitem{Maru:2022bhr}
N.~Maru and A.~Okawa, \emph{{Cosmological Collider Signals of Non-Gaussianity
  from Higgs boson in GUT}},
  \href{https://arxiv.org/abs/2206.06651}{{\ttfamily 2206.06651}}.

\bibitem{Seery:2006vu}
D.~Seery, J.E.~Lidsey and M.S.~Sloth, \emph{{The inflationary trispectrum}},
  \href{https://doi.org/10.1088/1475-7516/2007/01/027}{\emph{JCAP} {\bfseries
  01} (2007) 027} [\href{https://arxiv.org/abs/astro-ph/0610210}{{\ttfamily
  astro-ph/0610210}}].

\bibitem{Arroja:2008ga}
F.~Arroja and K.~Koyama, \emph{{Non-gaussianity from the trispectrum in general
  single field inflation}},
  \href{https://doi.org/10.1103/PhysRevD.77.083517}{\emph{Phys. Rev. D}
  {\bfseries 77} (2008) 083517}
  [\href{https://arxiv.org/abs/0802.1167}{{\ttfamily 0802.1167}}].

\bibitem{Chen:2006dfn}
X.~Chen, M.-x.~Huang and G.~Shiu, \emph{{The Inflationary Trispectrum for
  Models with Large Non-Gaussianities}},
  \href{https://doi.org/10.1103/PhysRevD.74.121301}{\emph{Phys. Rev. D}
  {\bfseries 74} (2006) 121301}
  [\href{https://arxiv.org/abs/hep-th/0610235}{{\ttfamily hep-th/0610235}}].

\bibitem{Seery:2008ax}
D.~Seery, M.S.~Sloth and F.~Vernizzi, \emph{{Inflationary trispectrum from
  graviton exchange}},
  \href{https://doi.org/10.1088/1475-7516/2009/03/018}{\emph{JCAP} {\bfseries
  03} (2009) 018} [\href{https://arxiv.org/abs/0811.3934}{{\ttfamily
  0811.3934}}].

\bibitem{Domcke:2019mnd}
V.~Domcke, B.~von Harling, E.~Morgante and K.~Mukaida, \emph{{Baryogenesis from
  axion inflation}},
  \href{https://doi.org/10.1088/1475-7516/2019/10/032}{\emph{JCAP} {\bfseries
  10} (2019) 032} [\href{https://arxiv.org/abs/1905.13318}{{\ttfamily
  1905.13318}}].

\bibitem{Adshead:2012kp}
P.~Adshead and M.~Wyman, \emph{{Chromo-Natural Inflation: Natural inflation on
  a steep potential with classical non-Abelian gauge fields}},
  \href{https://doi.org/10.1103/PhysRevLett.108.261302}{\emph{Phys. Rev. Lett.}
  {\bfseries 108} (2012) 261302}
  [\href{https://arxiv.org/abs/1202.2366}{{\ttfamily 1202.2366}}].

\bibitem{Adshead:2015pva}
P.~Adshead, J.T.~Giblin, T.R.~Scully and E.I.~Sfakianakis,
  \emph{{Gauge-preheating and the end of axion inflation}},
  \href{https://doi.org/10.1088/1475-7516/2015/12/034}{\emph{JCAP} {\bfseries
  12} (2015) 034} [\href{https://arxiv.org/abs/1502.06506}{{\ttfamily
  1502.06506}}].

\bibitem{Adshead:2016iae}
P.~Adshead, J.T.~Giblin, T.R.~Scully and E.I.~Sfakianakis,
  \emph{{Magnetogenesis from axion inflation}},
  \href{https://doi.org/10.1088/1475-7516/2016/10/039}{\emph{JCAP} {\bfseries
  10} (2016) 039} [\href{https://arxiv.org/abs/1606.08474}{{\ttfamily
  1606.08474}}].

\bibitem{Adshead:2018doq}
P.~Adshead, J.T.~Giblin and Z.J.~Weiner, \emph{{Gravitational waves from gauge
  preheating}}, \href{https://doi.org/10.1103/PhysRevD.98.043525}{\emph{Phys.
  Rev. D} {\bfseries 98} (2018) 043525}
  [\href{https://arxiv.org/abs/1805.04550}{{\ttfamily 1805.04550}}].

\bibitem{Adshead:2019lbr}
P.~Adshead, J.T.~Giblin, M.~Pieroni and Z.J.~Weiner, \emph{{Constraining axion
  inflation with gravitational waves from preheating}},
  \href{https://doi.org/10.1103/PhysRevD.101.083534}{\emph{Phys. Rev. D}
  {\bfseries 101} (2020) 083534}
  [\href{https://arxiv.org/abs/1909.12842}{{\ttfamily 1909.12842}}].

\bibitem{Adshead:2019igv}
P.~Adshead, J.T.~Giblin, M.~Pieroni and Z.J.~Weiner, \emph{{Constraining Axion
  Inflation with Gravitational Waves across 29 Decades in Frequency}},
  \href{https://doi.org/10.1103/PhysRevLett.124.171301}{\emph{Phys. Rev. Lett.}
  {\bfseries 124} (2020) 171301}
  [\href{https://arxiv.org/abs/1909.12843}{{\ttfamily 1909.12843}}].

\bibitem{Freese:1990rb}
K.~Freese, J.A.~Frieman and A.V.~Olinto, \emph{{Natural inflation with pseudo -
  Nambu-Goldstone bosons}},
  \href{https://doi.org/10.1103/PhysRevLett.65.3233}{\emph{Phys. Rev. Lett.}
  {\bfseries 65} (1990) 3233}.

\bibitem{Silverstein:2008sg}
E.~Silverstein and A.~Westphal, \emph{{Monodromy in the CMB: Gravity Waves and
  String Inflation}},
  \href{https://doi.org/10.1103/PhysRevD.78.106003}{\emph{Phys. Rev. D}
  {\bfseries 78} (2008) 106003}
  [\href{https://arxiv.org/abs/0803.3085}{{\ttfamily 0803.3085}}].

\bibitem{McAllister:2008hb}
L.~McAllister, E.~Silverstein and A.~Westphal, \emph{{Gravity Waves and Linear
  Inflation from Axion Monodromy}},
  \href{https://doi.org/10.1103/PhysRevD.82.046003}{\emph{Phys. Rev. D}
  {\bfseries 82} (2010) 046003}
  [\href{https://arxiv.org/abs/0808.0706}{{\ttfamily 0808.0706}}].

\bibitem{Kim:2004rp}
J.E.~Kim, H.P.~Nilles and M.~Peloso, \emph{{Completing natural inflation}},
  \href{https://doi.org/10.1088/1475-7516/2005/01/005}{\emph{JCAP} {\bfseries
  01} (2005) 005} [\href{https://arxiv.org/abs/hep-ph/0409138}{{\ttfamily
  hep-ph/0409138}}].

\bibitem{Berg:2009tg}
M.~Berg, E.~Pajer and S.~Sjors, \emph{{Dante's Inferno}},
  \href{https://doi.org/10.1103/PhysRevD.81.103535}{\emph{Phys. Rev. D}
  {\bfseries 81} (2010) 103535}
  [\href{https://arxiv.org/abs/0912.1341}{{\ttfamily 0912.1341}}].

\bibitem{Dimopoulos:2005ac}
S.~Dimopoulos, S.~Kachru, J.~McGreevy and J.G.~Wacker, \emph{{N-flation}},
  \href{https://doi.org/10.1088/1475-7516/2008/08/003}{\emph{JCAP} {\bfseries
  08} (2008) 003} [\href{https://arxiv.org/abs/hep-th/0507205}{{\ttfamily
  hep-th/0507205}}].

\bibitem{Pajer:2013fsa}
E.~Pajer and M.~Peloso, \emph{{A review of Axion Inflation in the era of
  Planck}}, \href{https://doi.org/10.1088/0264-9381/30/21/214002}{\emph{Class.
  Quant. Grav.} {\bfseries 30} (2013) 214002}
  [\href{https://arxiv.org/abs/1305.3557}{{\ttfamily 1305.3557}}].

\bibitem{Anber:2006xt}
M.M.~Anber and L.~Sorbo, \emph{{N-flationary magnetic fields}},
  \href{https://doi.org/10.1088/1475-7516/2006/10/018}{\emph{JCAP} {\bfseries
  10} (2006) 018} [\href{https://arxiv.org/abs/astro-ph/0606534}{{\ttfamily
  astro-ph/0606534}}].

\bibitem{Anber:2009ua}
M.M.~Anber and L.~Sorbo, \emph{{Naturally inflating on steep potentials through
  electromagnetic dissipation}},
  \href{https://doi.org/10.1103/PhysRevD.81.043534}{\emph{Phys. Rev. D}
  {\bfseries 81} (2010) 043534}
  [\href{https://arxiv.org/abs/0908.4089}{{\ttfamily 0908.4089}}].

\bibitem{Cook:2011hg}
J.L.~Cook and L.~Sorbo, \emph{{Particle production during inflation and
  gravitational waves detectable by ground-based interferometers}},
  \href{https://doi.org/10.1103/PhysRevD.85.023534}{\emph{Phys. Rev. D}
  {\bfseries 85} (2012) 023534}
  [\href{https://arxiv.org/abs/1109.0022}{{\ttfamily 1109.0022}}].

\bibitem{Barnaby:2010vf}
N.~Barnaby and M.~Peloso, \emph{{Large Nongaussianity in Axion Inflation}},
  \href{https://doi.org/10.1103/PhysRevLett.106.181301}{\emph{Phys. Rev. Lett.}
  {\bfseries 106} (2011) 181301}
  [\href{https://arxiv.org/abs/1011.1500}{{\ttfamily 1011.1500}}].

\bibitem{Barnaby:2011qe}
N.~Barnaby, E.~Pajer and M.~Peloso, \emph{{Gauge Field Production in Axion
  Inflation: Consequences for Monodromy, non-Gaussianity in the CMB, and
  Gravitational Waves at Interferometers}},
  \href{https://doi.org/10.1103/PhysRevD.85.023525}{\emph{Phys. Rev. D}
  {\bfseries 85} (2012) 023525}
  [\href{https://arxiv.org/abs/1110.3327}{{\ttfamily 1110.3327}}].

\bibitem{Barnaby:2011vw}
N.~Barnaby, R.~Namba and M.~Peloso, \emph{{Phenomenology of a Pseudo-Scalar
  Inflaton: Naturally Large Nongaussianity}},
  \href{https://doi.org/10.1088/1475-7516/2011/04/009}{\emph{JCAP} {\bfseries
  04} (2011) 009} [\href{https://arxiv.org/abs/1102.4333}{{\ttfamily
  1102.4333}}].

\bibitem{Meerburg:2012id}
P.D.~Meerburg and E.~Pajer, \emph{{Observational Constraints on Gauge Field
  Production in Axion Inflation}},
  \href{https://doi.org/10.1088/1475-7516/2013/02/017}{\emph{JCAP} {\bfseries
  02} (2013) 017} [\href{https://arxiv.org/abs/1203.6076}{{\ttfamily
  1203.6076}}].

\bibitem{Anber:2012du}
M.M.~Anber and L.~Sorbo, \emph{{Non-Gaussianities and chiral gravitational
  waves in natural steep inflation}},
  \href{https://doi.org/10.1103/PhysRevD.85.123537}{\emph{Phys. Rev. D}
  {\bfseries 85} (2012) 123537}
  [\href{https://arxiv.org/abs/1203.5849}{{\ttfamily 1203.5849}}].

\bibitem{Linde:2012bt}
A.~Linde, S.~Mooij and E.~Pajer, \emph{{Gauge field production in supergravity
  inflation: Local non-Gaussianity and primordial black holes}},
  \href{https://doi.org/10.1103/PhysRevD.87.103506}{\emph{Phys. Rev. D}
  {\bfseries 87} (2013) 103506}
  [\href{https://arxiv.org/abs/1212.1693}{{\ttfamily 1212.1693}}].

\bibitem{Cheng:2015oqa}
S.-L.~Cheng, W.~Lee and K.-W.~Ng, \emph{{Numerical study of pseudoscalar
  inflation with an axion-gauge field coupling}},
  \href{https://doi.org/10.1103/PhysRevD.93.063510}{\emph{Phys. Rev. D}
  {\bfseries 93} (2016) 063510}
  [\href{https://arxiv.org/abs/1508.00251}{{\ttfamily 1508.00251}}].

\bibitem{Garcia-Bellido:2016dkw}
J.~Garcia-Bellido, M.~Peloso and C.~Unal, \emph{{Gravitational waves at
  interferometer scales and primordial black holes in axion inflation}},
  \href{https://doi.org/10.1088/1475-7516/2016/12/031}{\emph{JCAP} {\bfseries
  12} (2016) 031} [\href{https://arxiv.org/abs/1610.03763}{{\ttfamily
  1610.03763}}].

\bibitem{Domcke:2016bkh}
V.~Domcke, M.~Pieroni and P.~Bin\'etruy, \emph{{Primordial gravitational waves
  for universality classes of pseudoscalar inflation}},
  \href{https://doi.org/10.1088/1475-7516/2016/06/031}{\emph{JCAP} {\bfseries
  06} (2016) 031} [\href{https://arxiv.org/abs/1603.01287}{{\ttfamily
  1603.01287}}].

\bibitem{Peloso:2016gqs}
M.~Peloso, L.~Sorbo and C.~Unal, \emph{{Rolling axions during inflation:
  perturbativity and signatures}},
  \href{https://doi.org/10.1088/1475-7516/2016/09/001}{\emph{JCAP} {\bfseries
  09} (2016) 001} [\href{https://arxiv.org/abs/1606.00459}{{\ttfamily
  1606.00459}}].

\bibitem{Domcke:2018eki}
V.~Domcke and K.~Mukaida, \emph{{Gauge Field and Fermion Production during
  Axion Inflation}},
  \href{https://doi.org/10.1088/1475-7516/2018/11/020}{\emph{JCAP} {\bfseries
  11} (2018) 020} [\href{https://arxiv.org/abs/1806.08769}{{\ttfamily
  1806.08769}}].

\bibitem{Cuissa:2018oiw}
J.R.C.~Cuissa and D.G.~Figueroa, \emph{{Lattice formulation of axion inflation.
  Application to preheating}},
  \href{https://doi.org/10.1088/1475-7516/2019/06/002}{\emph{JCAP} {\bfseries
  06} (2019) 002} [\href{https://arxiv.org/abs/1812.03132}{{\ttfamily
  1812.03132}}].

\bibitem{Sorbo:2011rz}
L.~Sorbo, \emph{{Parity violation in the Cosmic Microwave Background from a
  pseudoscalar inflaton}},
  \href{https://doi.org/10.1088/1475-7516/2011/06/003}{\emph{JCAP} {\bfseries
  06} (2011) 003} [\href{https://arxiv.org/abs/1101.1525}{{\ttfamily
  1101.1525}}].

\bibitem{Odintsov:2022hxu}
S.D.~Odintsov and V.K.~Oikonomou, \emph{{Chirality of gravitational waves in
  Chern-Simons f(R) gravity cosmology}},
  \href{https://doi.org/10.1103/PhysRevD.105.104054}{\emph{Phys. Rev. D}
  {\bfseries 105} (2022) 104054}
  [\href{https://arxiv.org/abs/2205.07304}{{\ttfamily 2205.07304}}].

\bibitem{Nojiri:2020pqr}
S.~Nojiri, S.D.~Odintsov, V.K.~Oikonomou and A.A.~Popov, \emph{{Propagation of
  gravitational waves in Chern\textendash{}Simons axion $F(R)$ gravity}},
  \href{https://doi.org/10.1016/j.dark.2020.100514}{\emph{Phys. Dark Univ.}
  {\bfseries 28} (2020) 100514}
  [\href{https://arxiv.org/abs/2002.10402}{{\ttfamily 2002.10402}}].

\bibitem{Nojiri:2019nar}
S.~Nojiri, S.D.~Odintsov, V.K.~Oikonomou and A.A.~Popov, \emph{{Propagation of
  Gravitational Waves in Chern-Simons Axion Einstein Gravity}},
  \href{https://doi.org/10.1103/PhysRevD.100.084009}{\emph{Phys. Rev. D}
  {\bfseries 100} (2019) 084009}
  [\href{https://arxiv.org/abs/1909.01324}{{\ttfamily 1909.01324}}].

\bibitem{Odintsov:2019mlf}
S.D.~Odintsov and V.K.~Oikonomou, \emph{{$f(R)$ Gravity Inflation with
  String-Corrected Axion Dark Matter}},
  \href{https://doi.org/10.1103/PhysRevD.99.064049}{\emph{Phys. Rev. D}
  {\bfseries 99} (2019) 064049}
  [\href{https://arxiv.org/abs/1901.05363}{{\ttfamily 1901.05363}}].

\bibitem{Cho:2019plw}
H.-T.~Cho and K.-W.~Ng, \emph{{Electromagnetic coupling effects in natural
  inflation}}, \href{https://doi.org/10.1088/1361-6382/ab8600}{\emph{Class.
  Quant. Grav.} {\bfseries 37} (2020) 165011}
  [\href{https://arxiv.org/abs/1907.06204}{{\ttfamily 1907.06204}}].

\bibitem{Orlando:2020oko}
G.~Orlando, M.~Pieroni and A.~Ricciardone, \emph{{Measuring Parity Violation in
  the Stochastic Gravitational Wave Background with the LISA-Taiji network}},
  \href{https://doi.org/10.1088/1475-7516/2021/03/069}{\emph{JCAP} {\bfseries
  03} (2021) 069} [\href{https://arxiv.org/abs/2011.07059}{{\ttfamily
  2011.07059}}].

\bibitem{Cahn:2021ltp}
R.N.~Cahn, Z.~Slepian and J.~Hou, \emph{{A Test for Cosmological Parity
  Violation Using the 3D Distribution of Galaxies}},
  \href{https://arxiv.org/abs/2110.12004}{{\ttfamily 2110.12004}}.

\bibitem{Hou:2022wfj}
J.~Hou, Z.~Slepian and R.N.~Cahn, \emph{{Measurement of Parity-Odd Modes in the
  Large-Scale 4-Point Correlation Function of SDSS BOSS DR12 CMASS and LOWZ
  Galaxies}},  \href{https://arxiv.org/abs/2206.03625}{{\ttfamily 2206.03625}}.

\bibitem{Philcox:2022hkh}
O.H.E.~Philcox, \emph{{Probing parity violation with the four-point correlation
  function of BOSS galaxies}},
  \href{https://doi.org/10.1103/PhysRevD.106.063501}{\emph{Phys. Rev. D}
  {\bfseries 106} (2022) 063501}
  [\href{https://arxiv.org/abs/2206.04227}{{\ttfamily 2206.04227}}].

\bibitem{Shiraishi:2016mok}
M.~Shiraishi, \emph{{Parity violation in the CMB trispectrum from the scalar
  sector}}, \href{https://doi.org/10.1103/PhysRevD.94.083503}{\emph{Phys. Rev.
  D} {\bfseries 94} (2016) 083503}
  [\href{https://arxiv.org/abs/1608.00368}{{\ttfamily 1608.00368}}].

\bibitem{Liu:2019fag}
T.~Liu, X.~Tong, Y.~Wang and Z.-Z.~Xianyu, \emph{{Probing P and CP Violations
  on the Cosmological Collider}},
  \href{https://doi.org/10.1007/JHEP04(2020)189}{\emph{JHEP} {\bfseries 04}
  (2020) 189} [\href{https://arxiv.org/abs/1909.01819}{{\ttfamily
  1909.01819}}].

\bibitem{Cabass:2022rhr}
G.~Cabass, S.~Jazayeri, E.~Pajer and D.~Stefanyszyn, \emph{{Parity violation in
  the scalar trispectrum: no-go theorems and yes-go examples}},
  \href{https://arxiv.org/abs/2210.02907}{{\ttfamily 2210.02907}}.

\bibitem{Cabass:2022oap}
G.~Cabass, M.M.~Ivanov and O.H.E.~Philcox, \emph{{Colliding Ghosts:
  Constraining Inflation with the Parity-Odd Galaxy Four-Point Function}},
  \href{https://arxiv.org/abs/2210.16320}{{\ttfamily 2210.16320}}.

\bibitem{Niu:2022quw}
X.~Niu, M.H.~Rahat, K.~Srinivasan and W.~Xue, \emph{{Gravitational Wave Probes
  of Massive Gauge Bosons at the Cosmological Collider}},
  \href{https://arxiv.org/abs/2211.14331}{{\ttfamily 2211.14331}}.

\bibitem{Weinberg:2005vy}
S.~Weinberg, \emph{{Quantum contributions to cosmological correlations}},
  \href{https://doi.org/10.1103/PhysRevD.72.043514}{\emph{Phys. Rev. D}
  {\bfseries 72} (2005) 043514}
  [\href{https://arxiv.org/abs/hep-th/0506236}{{\ttfamily hep-th/0506236}}].

\bibitem{Maldacena:2002vr}
J.M.~Maldacena, \emph{{Non-Gaussian features of primordial fluctuations in
  single field inflationary models}},
  \href{https://doi.org/10.1088/1126-6708/2003/05/013}{\emph{JHEP} {\bfseries
  05} (2003) 013} [\href{https://arxiv.org/abs/astro-ph/0210603}{{\ttfamily
  astro-ph/0210603}}].

\bibitem{Bunn:1996py}
E.F.~Bunn, A.R.~Liddle and M.J.~White, \emph{{Four-year COBE normalization of
  inflationary cosmologies}},
  \href{https://doi.org/10.1103/PhysRevD.54.R5917}{\emph{Phys. Rev. D}
  {\bfseries 54} (1996) R5917}
  [\href{https://arxiv.org/abs/astro-ph/9607038}{{\ttfamily
  astro-ph/9607038}}].

\bibitem{WMAP:2010qai}
{\scshape WMAP} collaboration, \emph{{Seven-Year Wilkinson Microwave Anisotropy
  Probe (WMAP) Observations: Cosmological Interpretation}},
  \href{https://doi.org/10.1088/0067-0049/192/2/18}{\emph{Astrophys. J. Suppl.}
  {\bfseries 192} (2011) 18} [\href{https://arxiv.org/abs/1001.4538}{{\ttfamily
  1001.4538}}].

\bibitem{Planck:2019kim}
{\scshape Planck} collaboration, \emph{{Planck 2018 results. IX. Constraints on
  primordial non-Gaussianity}},
  \href{https://doi.org/10.1051/0004-6361/201935891}{\emph{Astron. Astrophys.}
  {\bfseries 641} (2020) A9}
  [\href{https://arxiv.org/abs/1905.05697}{{\ttfamily 1905.05697}}].

\bibitem{Marzouk:2022utf}
K.~Marzouk, A.~Lewis and J.~Carron, \emph{{Constraints on
  \ensuremath{\tau}$_{NL}$ from Planck temperature and polarization}},
  \href{https://doi.org/10.1088/1475-7516/2022/08/015}{\emph{JCAP} {\bfseries
  08} (2022) 015} [\href{https://arxiv.org/abs/2205.14408}{{\ttfamily
  2205.14408}}].

\end{thebibliography}\endgroup
\newpage
\bibliographystyle{JHEP}
\end{document}